\documentclass[preprint]{aastex}

\usepackage{aas_macros}
\usepackage{amsmath}
\usepackage{subfigure}
\usepackage{graphicx}
\usepackage{txfonts}
\usepackage{subfigure}
\usepackage{natbib}
\usepackage{graphicx,natbib,amssymb,color}
\usepackage[normalem]{ulem} 
\usepackage[latin1]{inputenc}
\usepackage{cancel}
\usepackage{hyperref}

\definecolor{leuchtblau}{rgb}{0.4,0.4,1}

\begin{document}
\title{Diffusion of energetic particles in turbulent MHD plasmas}

\author{M. Wisniewski\footnote{now at: Deutsche Accumotive}}
\affil{Lehrstuhl f\"ur Astronomie, Universit\"at W\"urzburg, Emil-Fischer-Str. 31, D-97074 W\"urzburg}
\author{F. Spanier}
\affil{Lehrstuhl f\"ur Astronomie, Universit\"at W\"urzburg, Emil-Fischer-Str. 31, D-97074 W\"urzburg}
\email{fspanier@astro.uni-wuerzburg.de}
\author{R. Kissmann}
\affil{Institut f\"ur Astro- und Teilchenphysik, Universit\"at Innsbruck}

\begin{abstract}
In this paper we investigate the transport of energetic particles in turbulent plasmas. A numerical approach is used to simulate the effect of the background plasma on the motion of energetic protons. The background plasma is in a dynamically turbulent state found from numerical MHD simulations, where we use parameters typical for the heliosphere. The implications for the transport parameters (i.e. pitch-angle diffusion coefficients and mean free path) are calculated and deviations from the quasi-linear theory are discussed.
\end{abstract}

\maketitle

\section{Introduction}

Energetic particles are ubiquitous in astrophysical and space
plasmas. From the start of research in this field it was clear
that the transport of these particles through the background plasma
 is not governed by collisions, but by interaction with large-scale fields
embedded in the plasma, in this context the resonant wave particle interaction
is of special importance.  In this paper the interaction of solar
energetic particles (SEP) with the solar wind plasma is studied by
means of a numerical model the turbulent fields by magnetohydrodynamic (MHD)
simulations.

The electromagnetic fields of the solar wind plasma and the ISM are highly
turbulent and are correlated in space and time. The respective
correlation functions are a key ingredient to the
Fokker-Planck-coefficients - the relevant parameters for the
description of particle transport.

Complicated correlation functions and also anisotropies provide a high complexity of this
phenomenon. Turbulence itself is a non-linear theory hardly accessible by analytical methods. Combined with the non-linear coupling of charged particles it is usually necessary to simplify the problem in an analytical treatment. Up to now important aspects of turbulence have been neglected and significant simplifications have been made  both in theoretical models and in the design of numerical simulations:

The first theoretical description of cosmic ray transport was done by
Jokipii in 1966, who derived the so-called quasilinear theory (QLT)
\citep[cf.][]{jokki}. The main result of this work is the
Fokker-Planck-Equation (FP equation) as the basic equation of cosmic
ray transport with its Fokker-Planck-coefficients. It describes the
change of the distribution function of the particles with time. One
limiting assumption of the QLT is that the guiding centers of the
particles are not disturbed by the turbulence. This assumption is inaccurate especially for
the strong turbulence regimes .

Furthermore the predictions for diffusion perpendicular to the background magnetic field on the basis of the QLT were found to be incorrect by \citet{shalchi05}. This has been improved by higher-order theories
which allowed for a disturbance of the guiding center motion. But even these new theories are valid only for deviations from the guiding center theory so they are still approximations of the exact motion.

Since the introduction of the QLT several of those nonlinear theories
have been introduced, e.g., the 'Field-Line-Random-Walk-Theory'
(FLRW)-Theory (see \citet{1995PhRvL..75.2136M}), the BAM-Theory (see
\citet{1997ApJ...485..655B}), the CC $\&$ RR-Theory (see
\citet{1978PhRvL..40...38R} ; \citet{1978NucFu..18..353S}), the
'nonlinear guiding center theory' (NLGC;
see \citet{2003ApJ...590L..53M}) and the 'weakly nonlinear theory'
(WNLT,see \citet{2004ApJ...616..617S}).

Especially the WNLT yields much better results than the QLT with its
correction terms for the guiding center motion which are calculated on
the basis of the QLT. The influence of the changed distribution
function on the turbulence, however, and also intermittency effects are
neglected and the correlation functions are approximated via Delta
functions.

Besides the  analytical approaches there have also been
some numerical investigations of cosmic ray diffusion.  Since the 1970s
several simulations of particle transport have
been investigated: \citet{1978PhFl...21..361K} and
\citet{1996NPGeo...3...66M} described turbulence as a one dimensional
static phenomenon. \citet{1999ApJ...520..204G} simulated particle
transport with 2500 particles in a static
\emph{Composite-Turbulence}, i.e. turbulent modes may be decomposed into modes parallel and perpendicular to the background field. \citet{2002ApJ...578L.117Q} modeled the
turbulence with an ensemble of magnetostatic modes in Fourier space
which extend over several magnitudes in wavenumber space.

One of the first models using dynamical turbulence in these
simulations is the one by \citet{2001A&A...376..667M} who implemented
the dynamical effects of plasma waves by superposing 768 waves moving
with their phase velocity parallel to the background field. The correlation in time is, however,
expressed incorrectly as the wave damping is neglected. This and the
neglect of the correct correlation functions in space due to the assumption of
independent wave modes have a direct influence on the resulting
Fokker-Planck coefficients as these coefficients depend directly on
the correlation functions. The correlation functions are an
important ingredient of a transport theory and should also play a role
in numerical investigations.

In this paper we use a fully dynamical turbulence model, which is
computed consistently by solving the MHD equations describing the
evolution of the plasma under consideration. This means that the
turbulence description underlying the analysis in the current paper
does not use any assumptions regarding the correlations functions of
the turbulent fluctuations -- instead they can be computed from the
turbulent fields resulting from solving the MHD system. This has to be
seen as an important generalisation of the models discussed above. The
only parameters determining the turbulence are the strength and the
configuration of the driving force used to sustain the turbulence.

The main subject of this paper, the solar energetic particles (SEP), are for several reasons an interesting topic. In the first place they provide the major part of the nonthermal energy in the heliosphere, although the single particles are far less energetic than galactic cosmic rays. Furthermore they show variability. While galactic cosmic rays stem from sources far beyond the diffusion length scale, SEPs still carry variability patterns from the dynamic Sun. The SEPs themselves can be divided into different classes by their source: There are SEPs from flare events, which are accelerated by wave particle interaction (ion cyclotron \citealt{1997ApJ...477..940R} or Alfv\'en waves \citealt{1993ApJ...412..386M,1995ApJ...452..912M}). And SEP from coronal mass ejections, which are accelerated in shock waves (\citealt{2008JASTP..70..467V} showed that in this context self-generated waves also provide efficient acceleration).

The different emission mechanisms of SEPs at the same time relate to different plasma backgrounds. Here we consider a compressibly driven plasma and an incompressibly driven plasma. Even though we may have incompressible driving, the solar wind plasma may not be regarded as fully incompressible.

As the example of self-generated turbulence shows, it is quite difficult to disentangle transport and acceleration especially for the case of expanding CME shock waves. It is therefore a better approach to study flare events in more detail as here the main process is always wave particle interaction, which can be studied with our numerical approach.

The aim of this work is to derive transport parameters in a simulation which resembles the physical background of SEPs. This includes on the one hand the effect of dynamics turbulence and on the other hand compressible MHD plasmas.

\subsection{Numerical Investigation}

For the numerical investigation of the transport of SEPs in an MHD
turbulent plasma we implemented a code for the simultaneous simulation
of MHD-turbulence and test particle motion. By test particles we mean
that in this case the particle motion is determined by the background
plasma but the background plasma is not influenced by the particles.
We simulate a driven self-consistent plasma turbulence on the basis of
the ideal MHD-equations and inject the particles as soon as the turbulence is
fully developed. The particles are scattered by the turbulent magnetic fields
of the MHD-plasma. From their trajectories we calculate the basic
transport parameters (see chapter \ref{cap:impcr} for details) as the
FP-coefficients
\begin{eqnarray}
  D_{\mu\mu} &=& \lim_{t\to\infty} \frac{(\Delta \mu)^2}{2\cdot t}\\
    D_{pp} &=& \lim_{t\to\infty} \frac{(\Delta p)^2}{2\cdot t}
\end{eqnarray}
and the parallel mean free path
\begin{eqnarray}
 \lambda_{\parallel}(v)=\frac{3}{8}v\int_{-1}^1 d\mu \frac{(1-\mu^2)^2}{D_{\mu\mu}}.
\end{eqnarray}
With our test particle simulations we are able to describe the exact
motion of the particle. Therefore we are not limited to approximations of
the guiding center motion. Apart from that we describe the correlation
functions correctly, as the turbulent fields are self-consistently
evaluated. In conclusion this is the first
self-consistent numerical determination of the
Fokker-Planck-coefficients for a dynamical MHD-plasma.

In this manuscript we investigate general characteristics of the transport process.
This study is done for the plasma conditions in the heliosphere as it was measured by
satellite experiments. The results for the transport parameters are compared
to actual measurements.

The structure of the paper is as follows: In section \ref{sec_numsim}
we introduce the mathematical description and the numerical scheme.   In
section \ref{sec_results2} the simulation results are compared to  experimental data of a flare event.

\section{Numerical Simulations}
\label{sec_numsim}
For the present study we used a code for the simultaneous
simulation of a turbulent background plasma (by solving the MHD equations) and
embedded charged particles via a test particles description.  We start out from an unperturbed
plasma with a homogeneous background magnetic field and drive the turbulence at each time step
via the injection of long wavelength random fluctuations. No sooner than until the turbulence saturates the test particles are injected into the plasma. They are accelerated by the Lorentz force
arising from the magnetic field within the plasma. The particles and
the fluid are evolved with the same time step and the position and velocity of the particles is stored
for the whole remaining simulation time. From this data the relevant coefficients
and parameters are calculated with independent analysis programs.

Although the background fields change only slightly on the timescales of the particle motion especially for high energetic particles, a much bigger time step for the evolution of the magnetic fields compared to the particles' time step has not turned out to be advantageous. We attempted to use a low order interpolation in time for magnetic and velocity fields needed for the particle acceleration. But the numerical errors, especially for lower particle energies turned out to be very high, as on their time scales the background fields do change quite a bit. These problems may be lifted by using higher order interpolations in time, but the higher memory use, makes this unattractive. In the light of this test case we refrained from using different time steps for particles and fields.

In the following paragraphs we present the specifics of the numerical methods for
the plasma and particle simulations as well as the turbulence driver.

\subsection{MHD equations}
The background plasma is described by the ideal MHD equations in a periodic domain, where the
system of equations is closed by an isothermal equation of state.
Since our analysis is a principle one all variables  are computed using non-dimensional,
normalised units. For the four independent normalisation
variables we use the length of
the simulation domain $L_\textrm{scal}$, the mass of the hydrogen atom $m_0$, a
typical number density $n_0$, and the temperature of the system $T_0$
which directly relates to the speed of sound $c_s$. The magnetic field is also normalised by the speed of sound with the normalisation constant $B_0 = \sqrt{\mu_0 \rho_0} c_s$.
The resulting set of normalised MHD-equations is the following:
\begin{align}
  \frac{\partial \rho}{\partial t}
  &=
  -\nabla \cdot \vec s  \\
  \frac{\partial \vec s}{\partial t}
  &=
  -\nabla \cdot \left( \frac{\vec s \vec s }{\rho} +    \left( p + \frac{B^2}{2} \right)1-\vec B \vec B \right)+\vec F
  \\
  \label{EqInductionHyper}
  \frac{\partial \vec B}{\partial t}
  &=
  -\nabla \cdot \left(\frac{\vec s \vec B -\vec B \vec s}{\rho} \right) \\
  p
  &= \frac{c_s ^2}{u_0 ^2}\rho \stackrel{(here)}{=} \rho
\label{iso}
\end{align}
Here $\rho$ is the mass density, $\vec s= \rho \vec v$ is the momentum
density, $\vec{B}$ indicates the magnetic induction
and $p$ is the thermal pressure. The isothermal equation of state --
Eq. (\ref{iso}) -- is simplified by our choice for
the normalisation constants. $u_0$ is the normalisation constant for the velocity, here chosen as the sound speed.

\subsection{Numerical plasma model}

For the numerical solution of
the MHD equations we use a code which employs several optimised
solvers for the corresponding subproblems. For the hyperbolic part of
the differential equations we use a semidiscrete central-upwind scheme
\citep[see][for further details]{KurganovNoellePetrova2001SJSC}. The
underlying finite-volume method utilises an approximate Riemann
solver, which does not require any characteristic decomposition. The
resulting scheme is, thus, highly efficient and very robust at the
same time.

Due to its finite volume character the scheme works on
cell-averages. Therefore, we need some reconstruction to compute local
values at the cell-interfaces from these cell-averages. For this we
employ the second order minmod-limiter, which also has the property to
be total variation diminishing \citep[TVD, for an explanation of this
limiter and further references please
see][]{KurganovNoellePetrova2001SJSC}.

In order to satisfy the divergence-constraint of the magnetic field we
use the so-called constrained transport method. For this scheme the
induction equation is cast into the conservative form
\mbox{Eq. \eqref{EqInductionHyper}} to be able to compute the
hyperbolic fluxes. These are then used to calculate local electric
fields $\vec{E}$, by which the induction equation is written in the
form:
\begin{equation}
  \frac{\partial \vec{B}}{\partial t}
  =
  - \nabla \times \vec{E}
\end{equation}
This equation is then discretised in a way so that the divergence of
the magnetic field vanishes in a finite volume sense\citep[see,
  e.g.,][]{LondrilloDelZanna2000}. For further details on the
constrained transport scheme please refer to the publications by
\citet{EvansHawley1988ApJ} and \citet{BalsaraSpicer1999JcP}.

Finally the time-integration is done via a third order Runge-Kutta
integrator, which is used in the TVD-form by
\citet{JiangShu1996JcP}. This algorithm not only yields high temporal
resolution but is also highly efficient regarding memory costs by
buffering just one additional field for each variable. Due to this
choice of the time-integration method in combination with the base
scheme for the hyperbolic conservation-equations the CFL number should
not be chosen much higher than 0.4. In our case we implemented an
adaptive time-step size, which keeps the time-step safely below this
limit \citep[For further details regarding the code please refer
  to][and references therin]{KissmannEtAl2008MNRAS,
  KissmannEtAl2009JCP}.

\subsection{Turbulence driver}

In this work the turbulent fields giving rise to the particle
diffusion are computed using the MHD equations as discussed above. The
only remaining ambiguity in this case is the method to drive the
turbulence, which will now be discussed. Such a driving force is
necessary because the fluctuations decay at the smallest scales
resolved by the numerical mesh. The underlying idea for such a driving
force is simply to provide a random energy input on the largest scales
of the numerical grid. By solving the MHD equations we will then end
up with a spectral \emph{energy range}, where we supply the driving
force, a spectral \emph{inertial range} and a \emph{dissipation} range
at the smallest scales.

\label{turb_driver}
To drive the fluctuations we define a driving function $f_k$ in Fourier space with
\begin{eqnarray}
f_k=s\cdot k^{-9/5}\exp(2\pi \imath p)
\end{eqnarray}
where ``$s$'' and ``$p$'' are random numbers between zero and one. $s$ obeys a Gaussian distribution, whereas $p$ is uniformly distributed. This function $f_k$ yields the Fourier coefficients of the velocity perturbation used for driving the fluctuations in our simulations. After transforming to configuration space $f_k \longrightarrow f_x$ we make the distinction between compressible and incompressible driving. For the former we interpret $f$ as a scalar potential $\vec{\delta v} =\nabla f_x$, whereas for the latter we employ $\vec{\delta v} =\nabla \times \vec f_x$.  As $\vec f_x$ has, thus, to be a vectorial function in the case of incompressible driving we define each component separately with different random numbers.  Each component of the wave-vector $\vec k$ in the driving spectrum ranges from 1 to $k_\textrm{limit}$ yielding a maximum absolute value of $\mid\vec k\mid\le \sqrt{3 k_\textrm{limit}^2}$.  If not stated otherwise we chose $k_\textrm{limit}=3$. By subtracting the mean momentum from the fluctuating input fields the net momentum change is zero. The energy input is normalised  in a way that the resulting spectrum after reaching convergence has the same amplitude as it is found by measurements (the magnetic amplitude is roughly $\delta B/B_0=4\times 10^{-3}$). With that we are able to gain meaningful results for  realistic turbulence amplitudes. The statistic average of the resulting driving spectrum is isotropic.

With the choice of compressible or incompressible driving we can discriminate between two different physical scenarios: Compressible fluctuations can be injected into the surrounding medium by CMEs  and thus drive the turbulence. Incompressible fluctuations can be caused by the cosmic rays
themselves which can generate and amplify Alfv\'en waves (see \citet{2007ApJ...658..622V}). Also cascading of large scale Alfv\'en waves from the sun does play a role.

\subsection{Analysis of field quantities}

\label{en_spec}To quantify anisotropic effect we finally define the compressible and incompressible energy spectra. Both are calculated in Fourier space on the basis of the simulated velocity fields via
\begin{align}
  P_{comp}(k)&\equiv\left\| \widehat{k}\cdot \vec{v}(\vec k)  \right\|^2/2 \\
  P_{incomp}(k)&\equiv\left\|  \widehat{k}\times \vec{v}(\vec k)  \right\|^2/2.
\end{align}
where $\widehat{k}$ is the normalised wave vector. These spectra are used as an indicator for the strength of Alfv\'en (incompressible) and magnetosonic (compressible) waves. The overall energy spectrum of the velocity
fields is
\begin{eqnarray}
  P_{K}(k)\equiv v^2(k)/2
\end{eqnarray}
where the sum of $P_{comp}(k)$ and $P_{incomp}(k)$ is $P_{K}(k)$.
Additionally we define the omnidirectional energy spectrum of the fluctuating magnetic fields via
\begin{eqnarray}
  P_{B}\equiv\delta B^2(k)/2 \overline {\rho},
\end{eqnarray}
which is normalised to the Alfv\'en velocity. $\bar\rho$ is the mean mass density.

\subsection{Modeling the energetic particle transport}
\label{cap:impcr}
Energetic particles are treated as test particles. They are injected randomly into the simulation domain, with a well defined velocity but random uniformly distributed angle to the local magnetic field for each particle. The random distribution in space is also equivalent to a random distribution in phase angle. The particles are protons which are influenced by the electromagnetic field via the Lorentz force
\begin{eqnarray}
\vec{F}= q \left(\vec E + \vec v_p \times \vec B \right)=q\left(- \vec v \times \vec B + \vec v_p \times \vec B \right)
\end{eqnarray}
where $\vec v$ is the fluid velocity and $\vec B$ is the magnetic field of the  plasma, $\vec{v}_p$ and $q$ are the velocity and the charge of the particles respectively.  The back-reaction of the particles on the plasma is neglected. To simulate the motion of a particle as accurately as possible it is important that the fields on the particle's trajectory are precisely known. Approximating of the fields at the position of a particle just using the constant cell average of the MHD fluid  each cell turned out to be much too inaccurate and results in unstable particle trajectories in otherwise stable test cases. Also the interpolation of the cell averages using a  linear spline failed even for simple tests. In the end a three dimensional cubic Spline proved to be a very good choice. For the time evolution we used again a third order Runge-Kutta algorithm because of the small memory costs with comparatively high temporal resolution.

For the calculation of the FP-coefficients a high number of test particles is necessary to sample an ensemble average of the turbulent fields. With such a high number of particles the average can be taken over several uncorrelated regions of the plasma.

The particle number has to be high enough to resolve the turbulence well enough that all turbulent structures have an statistically balanced influence on the simulated coefficients. For all test particles we store the pitch angle with respect to the local magnetic field and the momentum after several timesteps. With this $D_{\mu\mu}(\mu,p)$ of the individual particles can be calculated via
\begin{eqnarray}
D_{\mu\mu}(id,\mu,p)=\frac{(\mu_{end}-\mu_{begin})^{2}}{2\cdot \Delta t}.
\label{eq:dmm}
\end{eqnarray}
To obtain the statistical transport properties we compute the average for all particles. In particular we average over the particles within each bin of the $\mu-$ phase space.
\begin{eqnarray}
D_{\mu\mu}(\mu)=\sum_{\mu\, in\, \Delta \mu} D_{\mu\mu}(id,\mu,p)/N
\end{eqnarray}
Here N is the number of particles within the pitch angle bin under consideration.
With this the parallel mean free path which is also accessible to experiments can be calculated according to
\begin{eqnarray}
 \lambda_{\parallel}(v)=\frac{3}{8}L_{scal}\sum_{\mu=-1}^1 \Delta \mu \frac{v(1-\Delta \mu^2)^2}{D_{\mu\mu}(\mu)\cdot L_{scal}}.
\label{eq:mfp}
\end{eqnarray}
where $L_{scal}$ is the size of the simulation
domain.  In all simulations presented in this paper we
start out from randomly distributed particles with respect to the
pitch angle but all with the same momentum.

We also tested the Taylor-Green-Kubo formalism \citep{taylor1922,1951JChPh..19.1036G,1957JPSJ...12..570K} for the analysis of diffusion coefficients. As a matter of fact this method seems to result in erroneous results and very slow convergence of diffusion coefficients and mean free path.

\section{Results}
\label{sec_results2}

The code has been tested thoroughly for the case of single gyroresonances. This has been described  in \citet{2011ASTRA...7...21S} in great detail. The gyroresonance is the most crucial testcase for the derivation of wave-particle interactions. Here we will focus on application to the transport of charged protons in the heliosphere.

\subsection{Setup and numerical study}
The only available experimental data for the transport of energetic particles in space plasma are from direct satellite measurements of SEP events within the heliosphere and highenergetic Jovian electrons. Both analytical models and numerical simulations have to reproduce these results to prove their validity, which we want to do in the following for our present model.

\begin {figure}
\begin{center}
\setlength{\unitlength}{0.00045\textwidth}
\subfigure[]{
  \includegraphics[width=0.35\textwidth]{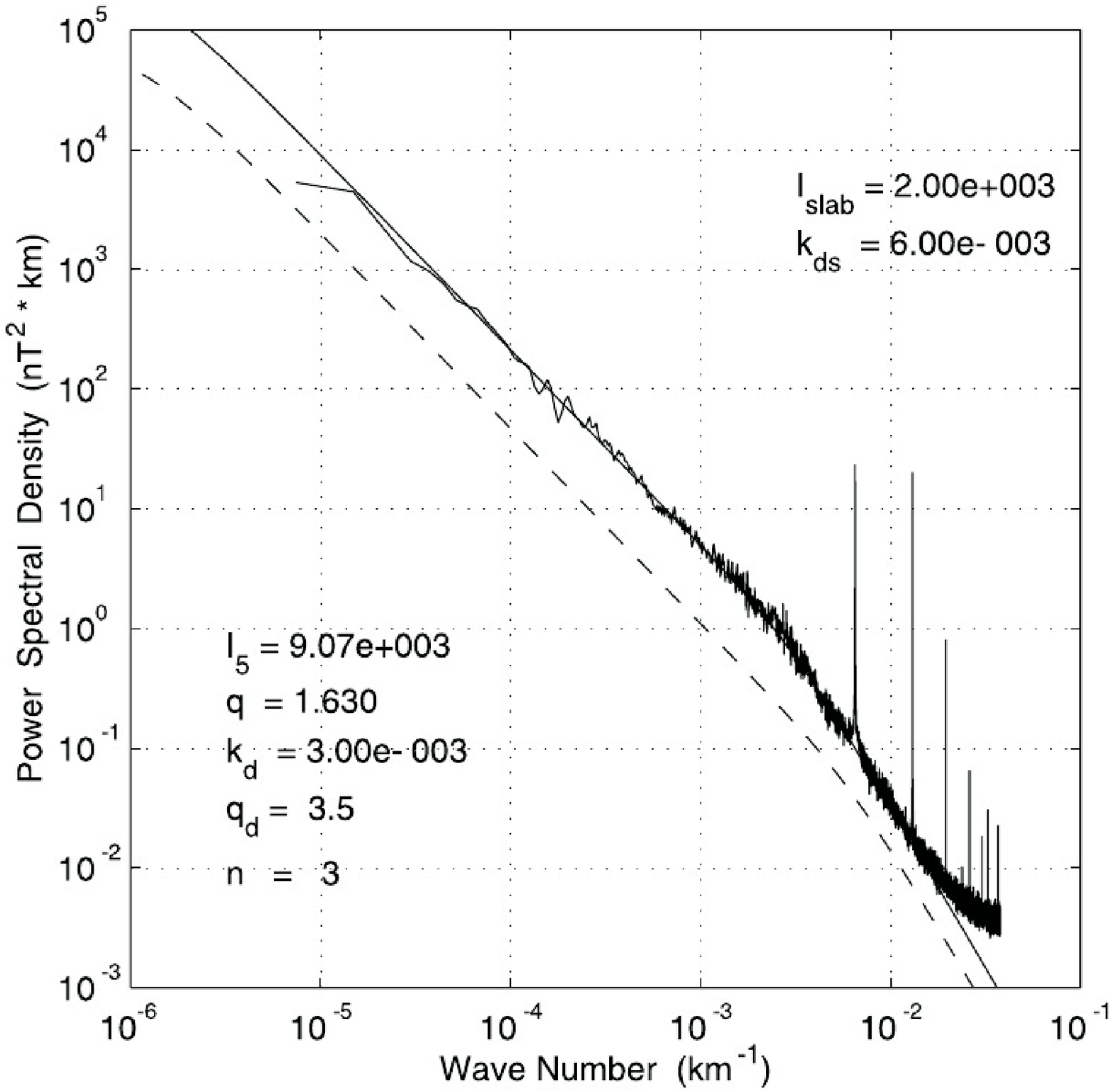}
  \label{fig:SpektrumExp}
}
\subfigure[]{
  \includegraphics[width=0.35\textwidth]{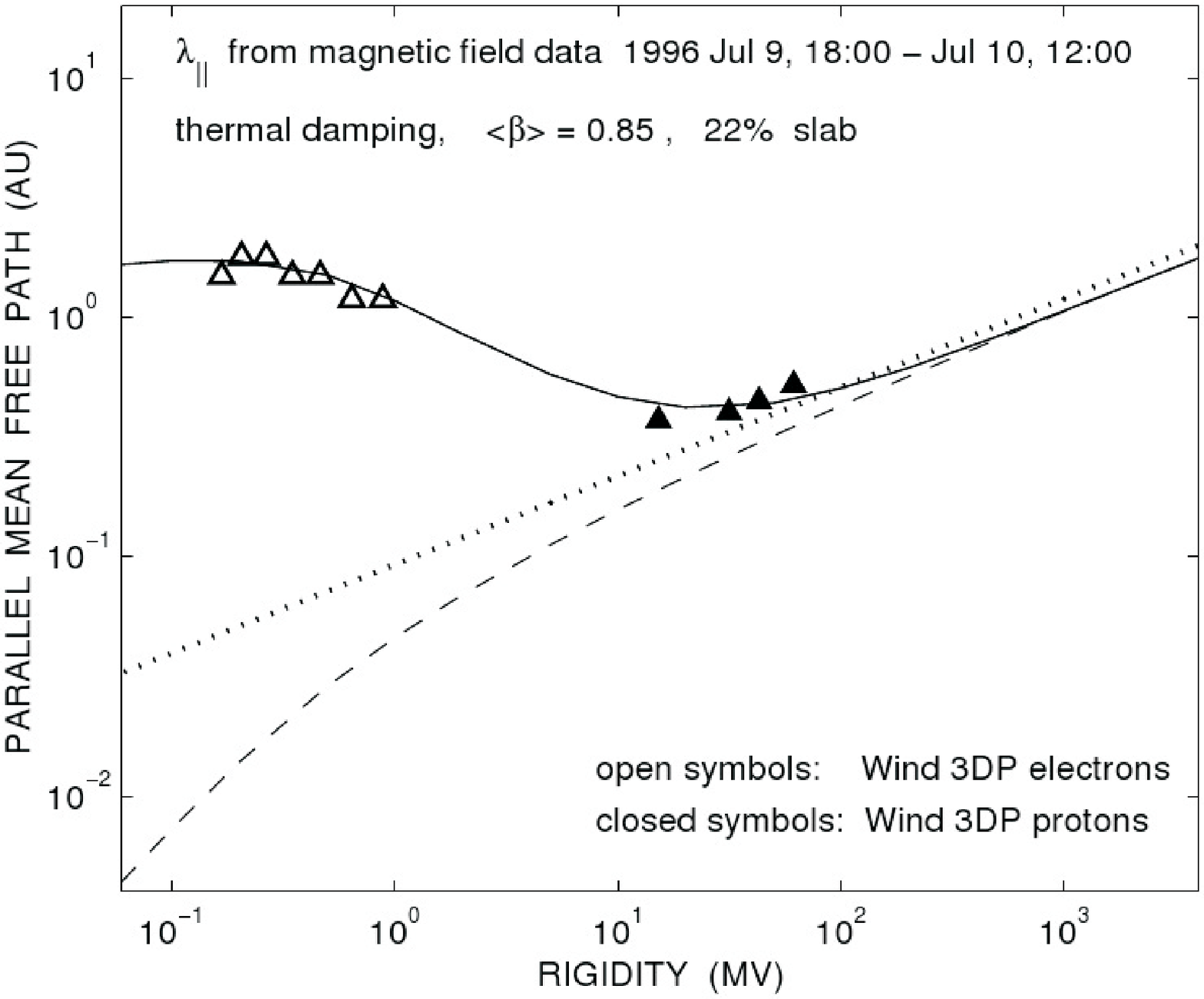}
  \label{fig:mfwlExp}
}
\end{center}
\caption{Fig. \ref{fig:SpektrumExp} shows the observed power spectrum of perpendicular fluctuation with a fit (solid line) and an estimate of the slab component (dashed line) for the observations in the period 1996 July 9 18:00 UT to July 10 12:00 UT. The slab component is estimated assuming that the power spectrum is defined by a 2D- and a slab-component. Using two transverse components the slab component can be estimated following \citet{bieber96}. Fig.
  \ref{fig:mfwlExp} shows the observed mean free path of Wind 3DP electron and proton data and predictions from the thermal damping model by \citet{2003ApJ...589.1027D}}
\end {figure}

Here we compare our simulation results to the measurements of the solar particles event for the period 1996 July 9 18:00 UT to July 10 12:00 UT by the Wind 3DP experiment, where the data is taken from \citet{2003ApJ...589.1027D}. The measured plasma parameters are: magnetic field strength \mbox{$B$=3.73 $\cdot 10^{-9}\,$ T}, particle density \mbox{n=5.27 $\cdot 10^6\, $m$^{-3}$}, temperature \mbox{$T$=59116 K}, Alfv\'en speed \mbox{$v_A=3542$ m/s} and \mbox{$\beta$=0.85.} Fig. \ref{fig:SpektrumExp} shows the magnetic field fluctuation spectrum measured by the Wind 3DP experiment. In Fig. \ref{fig:mfwlExp} the results for the parallel mean free path as a function of rigidity are shown, where the particles' rigidity is defined as $R=p/q \cdot c$. On the basis of those plasma parameters we simulated a spectrum, which has the same macroscopic observables, and injected particles within the measured energy range. We calculated their resulting mean free path as described in section \ref{cap:impcr}. These results are compared to the experimental data and give us an impression how SEP transport might work in the heliosphere.

\begin{figure}
\begin{center}
\setlength{\unitlength}{0.00045\textwidth}
\subfigure[]{
  \includegraphics[width=0.5\textwidth]{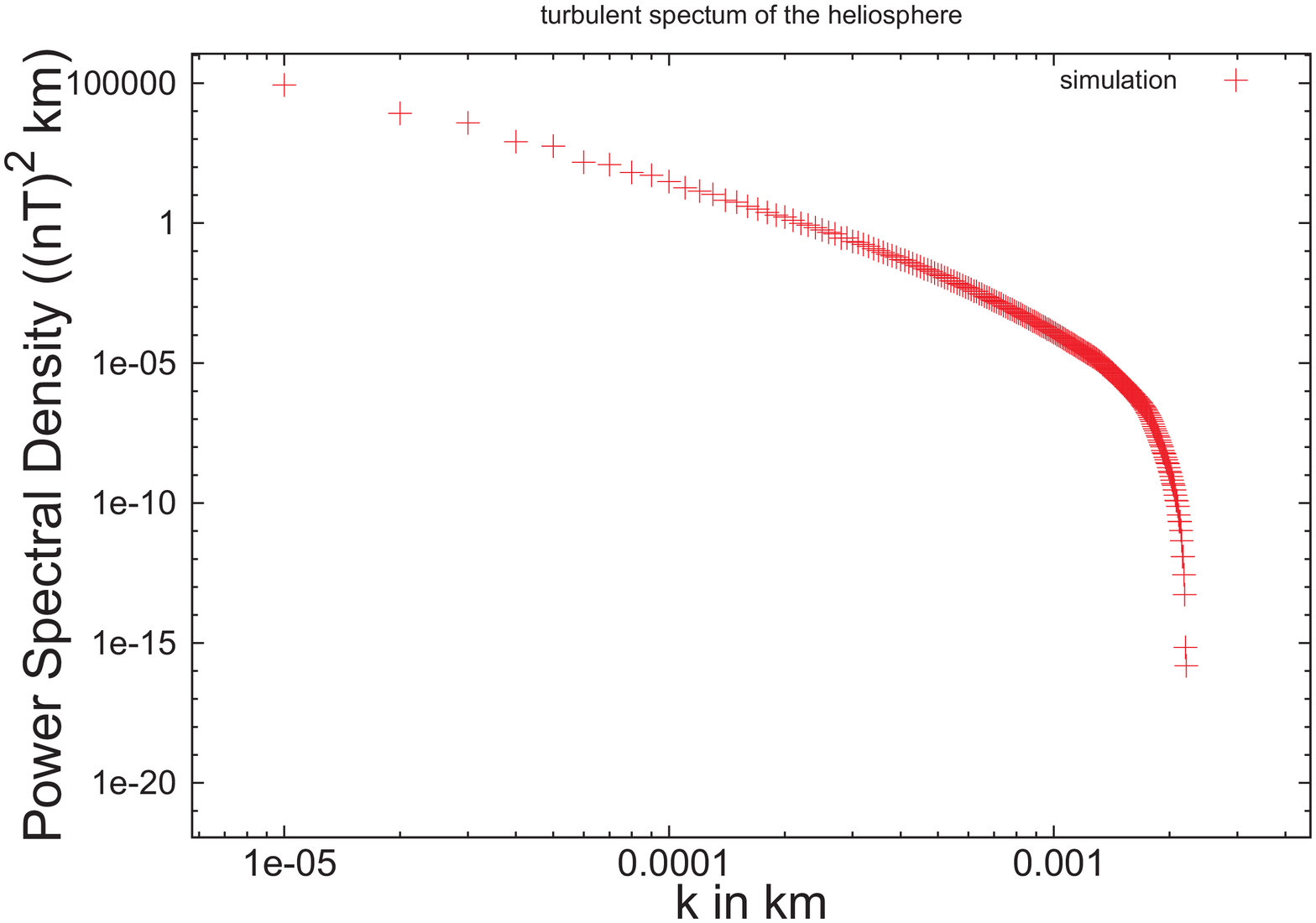}
  \label{fig:heli_spek}
}
\subfigure[]{
  \includegraphics[width=0.5\textwidth]{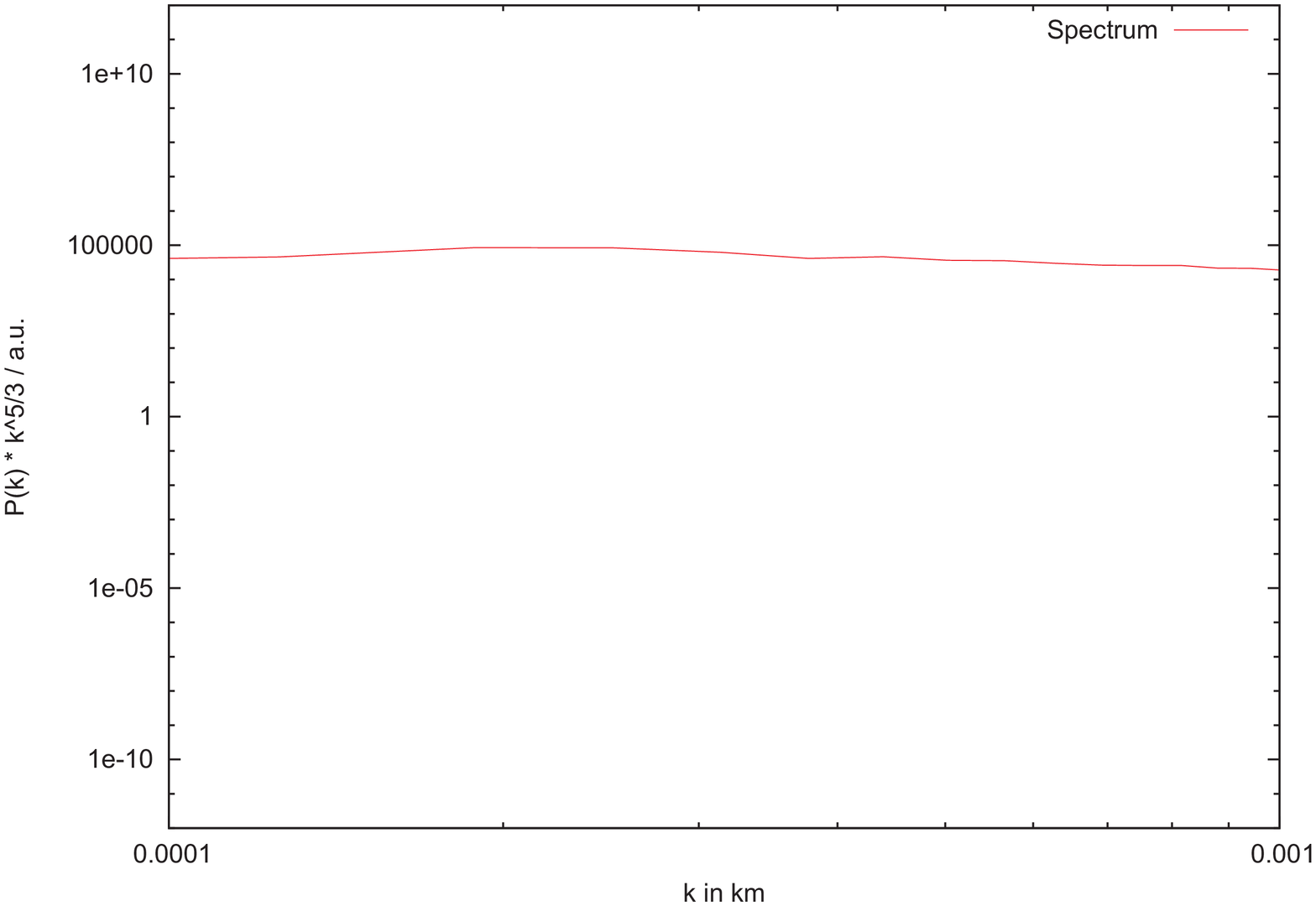}
  \label{fig:fuenfdreispek}
}
\end{center}
\caption{Simulated spectrum for our testparticle simulation for the heliospheric parameters which corresponds to the measured spectrum from the period 1996 July 9 18:00 UT to July 10 12:00 UT shown in \ref{fig:SpektrumExp}. It has to be noted, that the spectrum here is systemically higher than the observational spectrum. We follow the fit to the spectrum and show the total rather than the perpendicular or slab spectrum.\ref{fig:fuenfdreispek} additionally shows the spectrum multiplied with $k^{5/3}$ for the first decade.}
\end {figure}

Of course the present simulations are limited. If we have a closer look on the spectrum in Fig. \ref{fig:SpektrumExp} we see that the inertial range extends over four orders of magnitude. Just to be able to simulate the same extent of the inertial range would imply a grid resolution of more than $10^{4}$ gridpoints in each dimension. This is presently impossible even for the biggest supercomputers. Nevertheless it is possible to set up meaningful simulations as the particles of a certain energy interact mainly with those waves they are resonant with. Fig. \ref{fig:heli_spek} shows the simulated spectrum for a grid resolution of $256{^3}$. This spectrum contains all resonant modes for the simulated particles.

In general a comparison of the present spectrum  with simulations of \citet{2003ApJ...589.1027D} is difficult in many ways. Both spectra have in common some inertial range and a similar amplitude. But differences exist in the interpretation. The spectrum presented here consists of near-isotropic incompressible and compressible, whereas Dröge uses a slightly anisotropic incompressible spectrum. The advantage of using the present spectrum is its self-consistency. Even though this spectrum exceeds the amplitude of Dröge's model spectrum, perpendicular magnetic field fluctuations are of the same order in both spectra.

As the resonant wavenumber $\mid k_{res}\mid$ for a given particle energy and pitch angle depends on the propagation direction of the wave one has to be aware of the fact that several wave-modes could in principle interact resonantly with the particles. Additionally, we have to deal with a finite resonance width, so we do not only find an interaction at $k=k_{res}$ but also for slightly smaller and larger wavenumbers. It is of course crucial that all modes that contribute significantly to the acceleration of the particles are captured in the simulation. This question will now explicitly be analyzed for our simulations.

\subsubsection{Extent of the turbulent spectrum}

\begin {figure}
\begin{center}
\setlength{\unitlength}{0.00045\textwidth}
\subfigure[]{
  \includegraphics[width=0.5\textwidth]{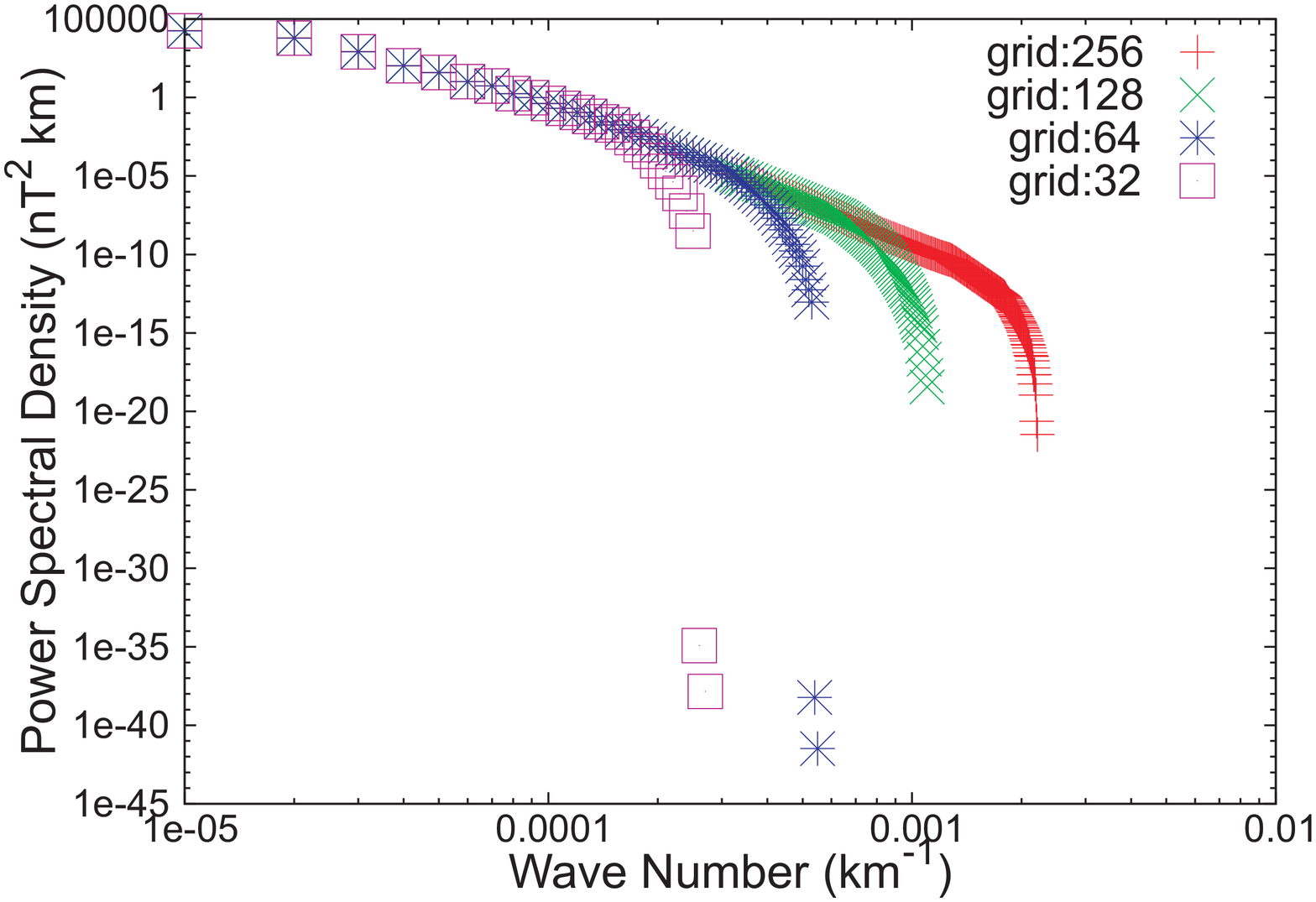}
  \label{fig:subfigFelder_PB_t50b10C2}
}
\subfigure[]{
  \includegraphics[width=0.5\textwidth]{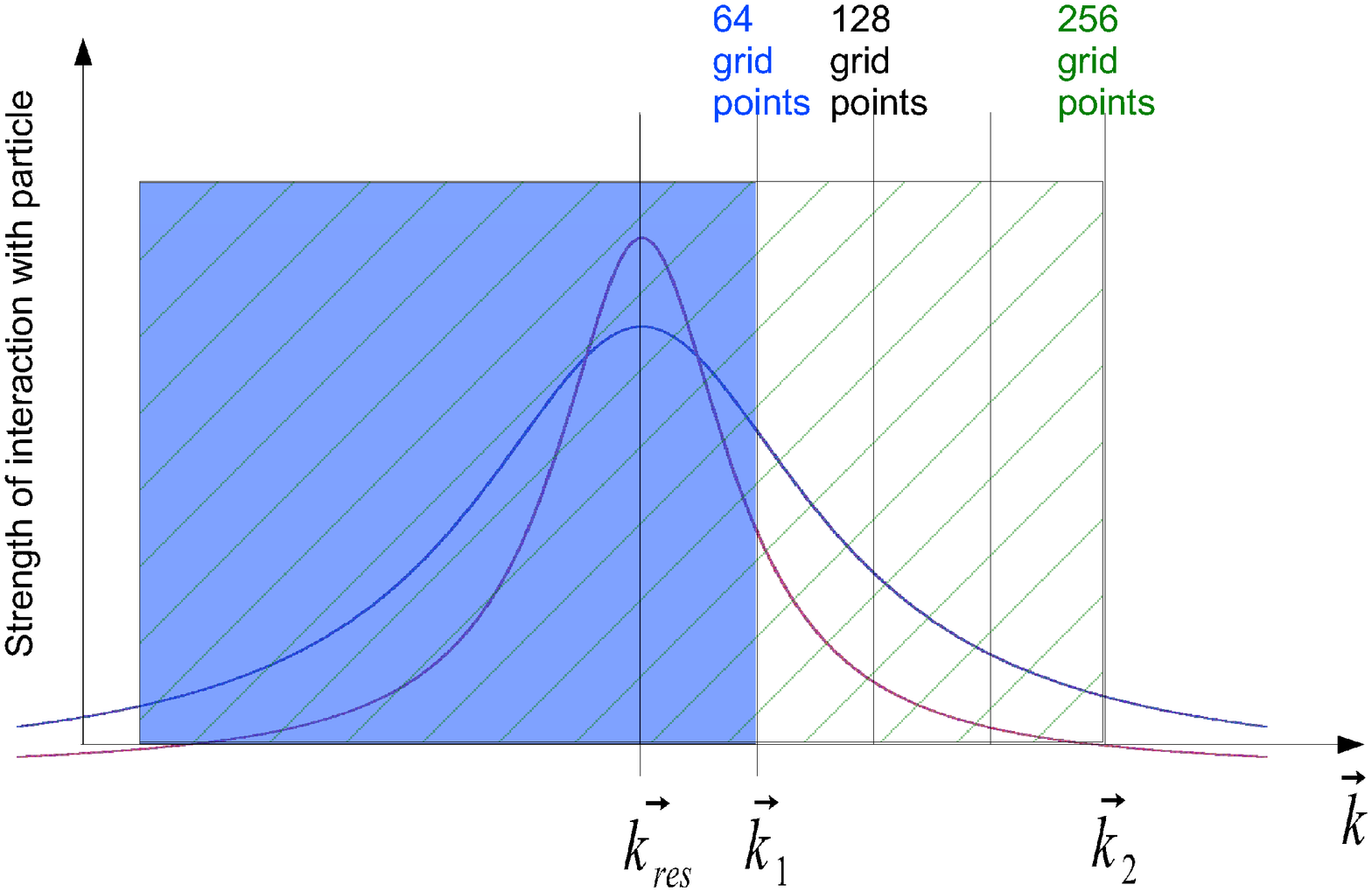}
  \label{fig:subfigFelder_PB_t50b10C2_bla}
}
\end{center}
\caption{ Fig. \ref{fig:subfigFelder_PB_t50b10C2} shows the spectra used for the convergence test for a $32^3$, $64^3$, $128^3$ and a $256^3$ grid.  These spectra have been computed by cutting the $256^3$ spectrum in Fourier space. Fig. \ref{fig:subfigFelder_PB_t50b10C2_bla} shows  how the strength of the wave particle interaction depends on the width of the simulated spectrum. The resonance width is unknown but exemplarily illustrated by the red and the blue curves. For higher grid resolution more modes can be resolved and contribute to the particle scattering. For the $64^3$ grid all modes up to $\vec k_{1}$ (blue area) whereas for the $256^3$ grid all modes up to $\vec k_{2}$ (green shaded area) can be resolved. The minimum grid resolution depends on the resonance width of the interaction and on the different $\vec k_{res}$ that can occur depending on the actual  physical parameters of the simulation.}
\label{fig:numerischerTestfall}
\end {figure}

The numerical particles interact with the MHD turbulent model-plasma. After about 10 s the parallel mean free path is computed from the particle trajectories. This is done for spatial grids with  $256^3$, $128^3$, $64^3$ and $32^3$ gridpoints respectively. For this testcase turbulence has converged, but the mean free path did not, so the absolute values of the mean free path are not important here. We are only interested in the relative effect of the different spatial resolutions. To generate the spectra on the smaller grids we didn't run new simulations but we cut the resulting spectrum on the $256^3$ grid in Fourierspace in a way, that the spectra correspond to simulations on a $128^3$ grid, a $64^3$ grid and a $32^3$ grid (cf. Fig. \ref{fig:subfigFelder_PB_t50b10C2}). The advantage of cutting the spectrum instead of simulating a new one on a smaller grid is that we can be sure that the same amount of energy is in each mode on the smaller grids with only those modes with the larger wavenumbers missing. With this method it is possible to determine whether the grid resolution is sufficient for the simulation of particles of a certain energy even though we are unable to incorporate the full extent of the measured turbulence spectrum. Test results are presented for protons with a rigidity of 17 MV.

\begin {figure}
\begin{center}
  \includegraphics[width=0.5\textwidth]{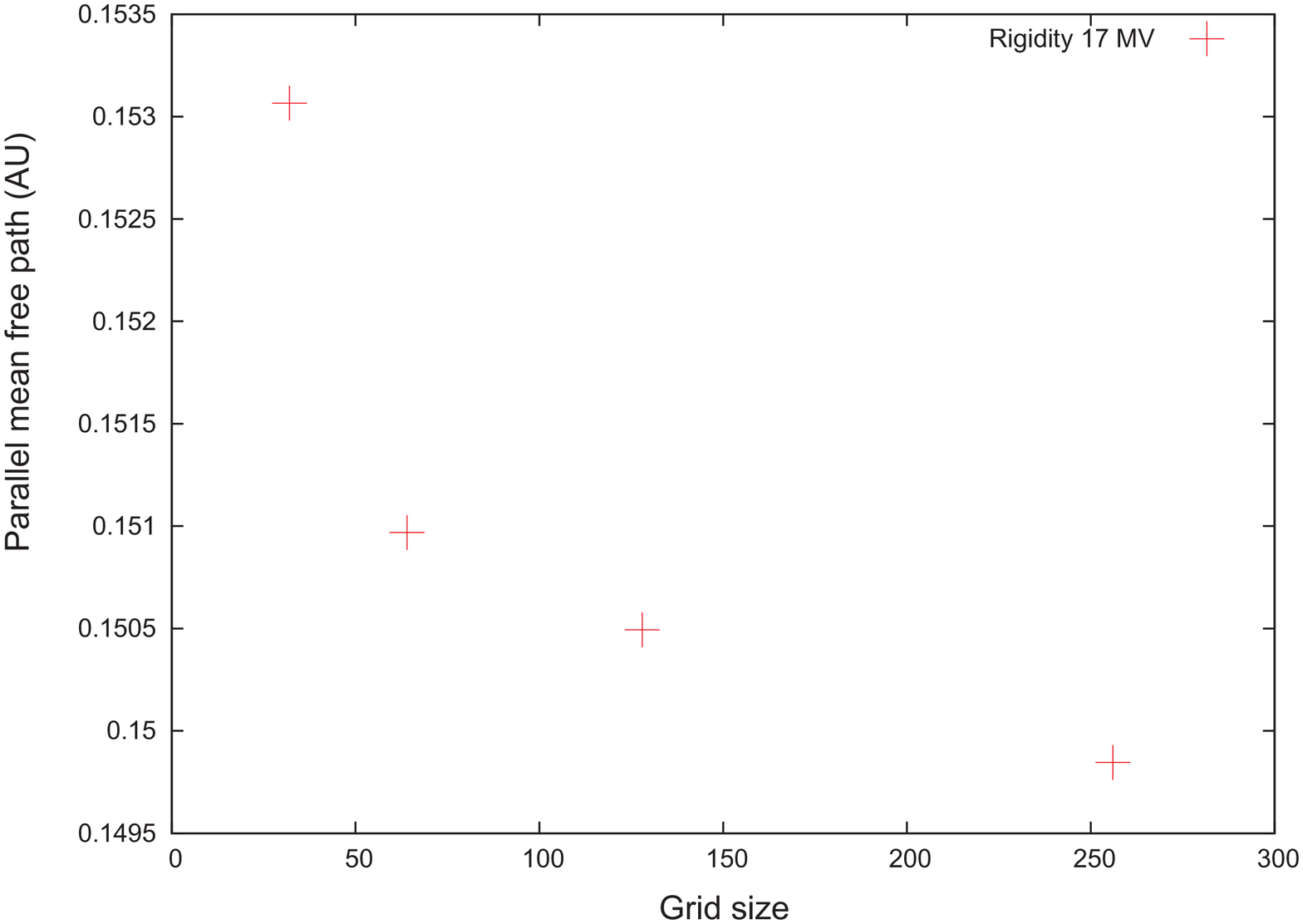}
\end{center}
\caption{Resulting parallel mean free path for a $32^3$, $64^3$, $128^3$ and $256^3$ grid. Convergence is obvious here.}
\label{fig:gitterstabil}
\end {figure}

The results are shown in Fig.  \ref{fig:gitterstabil}. As the mean free path does not depend on the grid resolution in this case one can draw the conclusion that higher wave modes do not influence the particle scattering drastically. This, however, does not necessarily allow the simulation of much smaller grids, as for stronger turbulence higher modes might have more influence and spectra tend to be even steeper.

\begin {figure}
\begin{center}
\setlength{\unitlength}{0.0003\textwidth}
\subfigure[PB]{
\begin{picture}(1000,837)(-100,-100)
\put(-60,330){\rotatebox{90}{$\ln k_{\perp}$}}
\put(360,-60){$\ln k_{\parallel}$}
\includegraphics[width=900\unitlength]{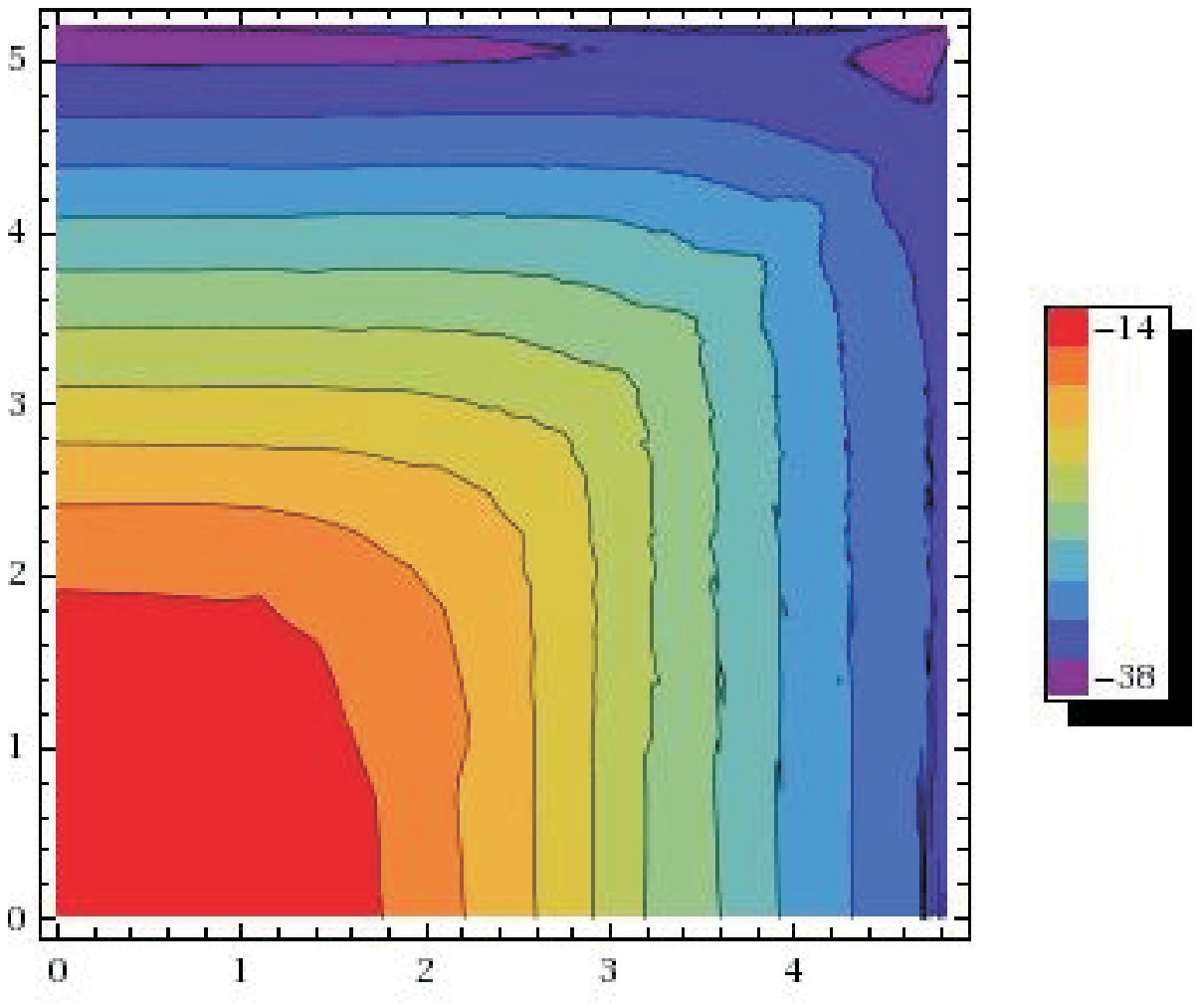}
\end{picture}
\label{PB}
}
\subfigure[PK]{
\begin{picture}(1000,837)(-100,-100)
\put(-60,330){\rotatebox{90}{$\ln k_{\perp}$}}
\put(360,-60){$\ln k_{\parallel}$}
    \includegraphics[width=900\unitlength]{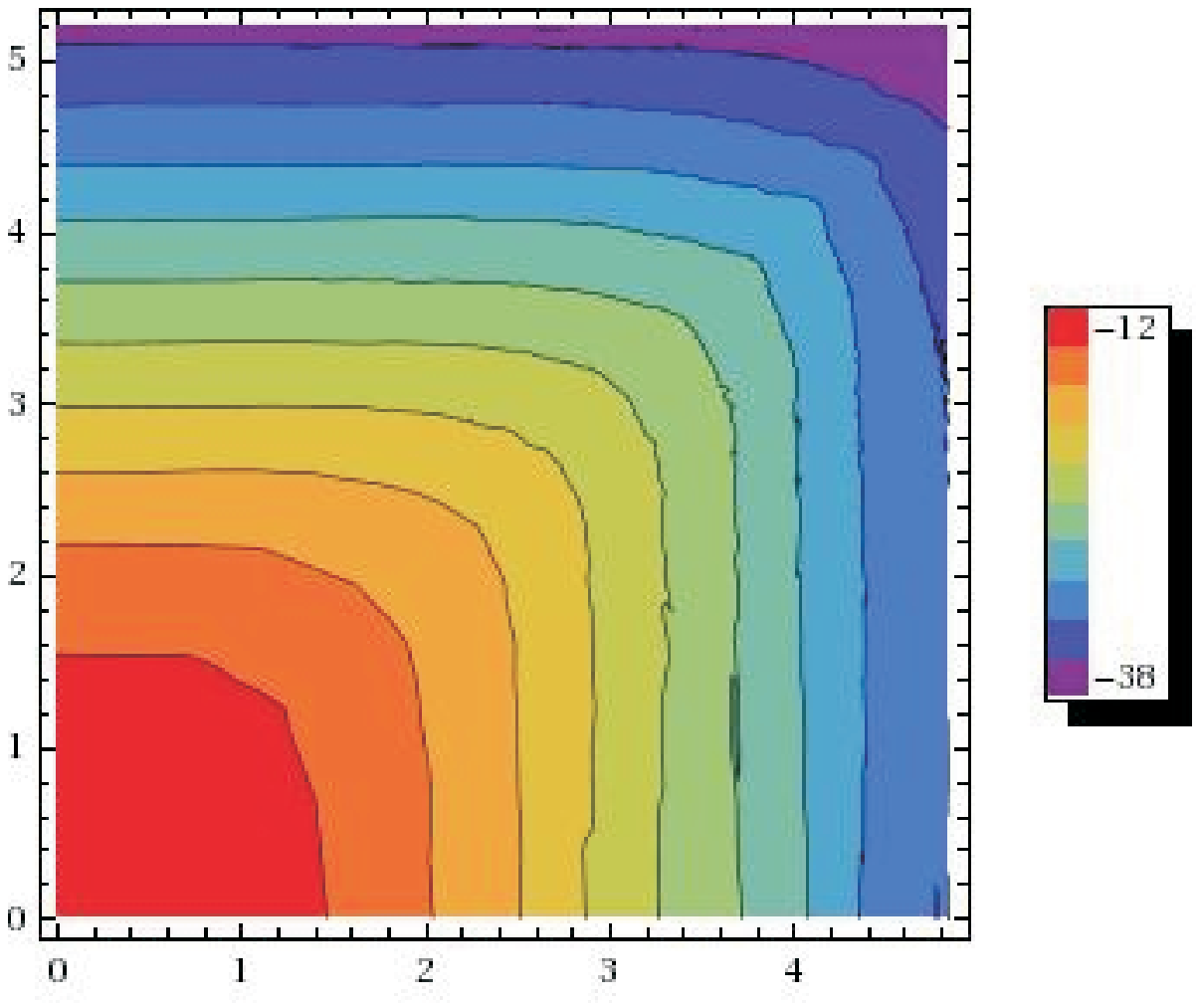}
  \end{picture}
  \label{PK}
}
\subfigure[PShear]{
\begin{picture}(1000,837)(-100,-100)
\put(-60,330){\rotatebox{90}{$\ln k_{\perp}$}}
\put(360,-60){$\ln k_{\parallel}$}
    \includegraphics[width=900\unitlength]{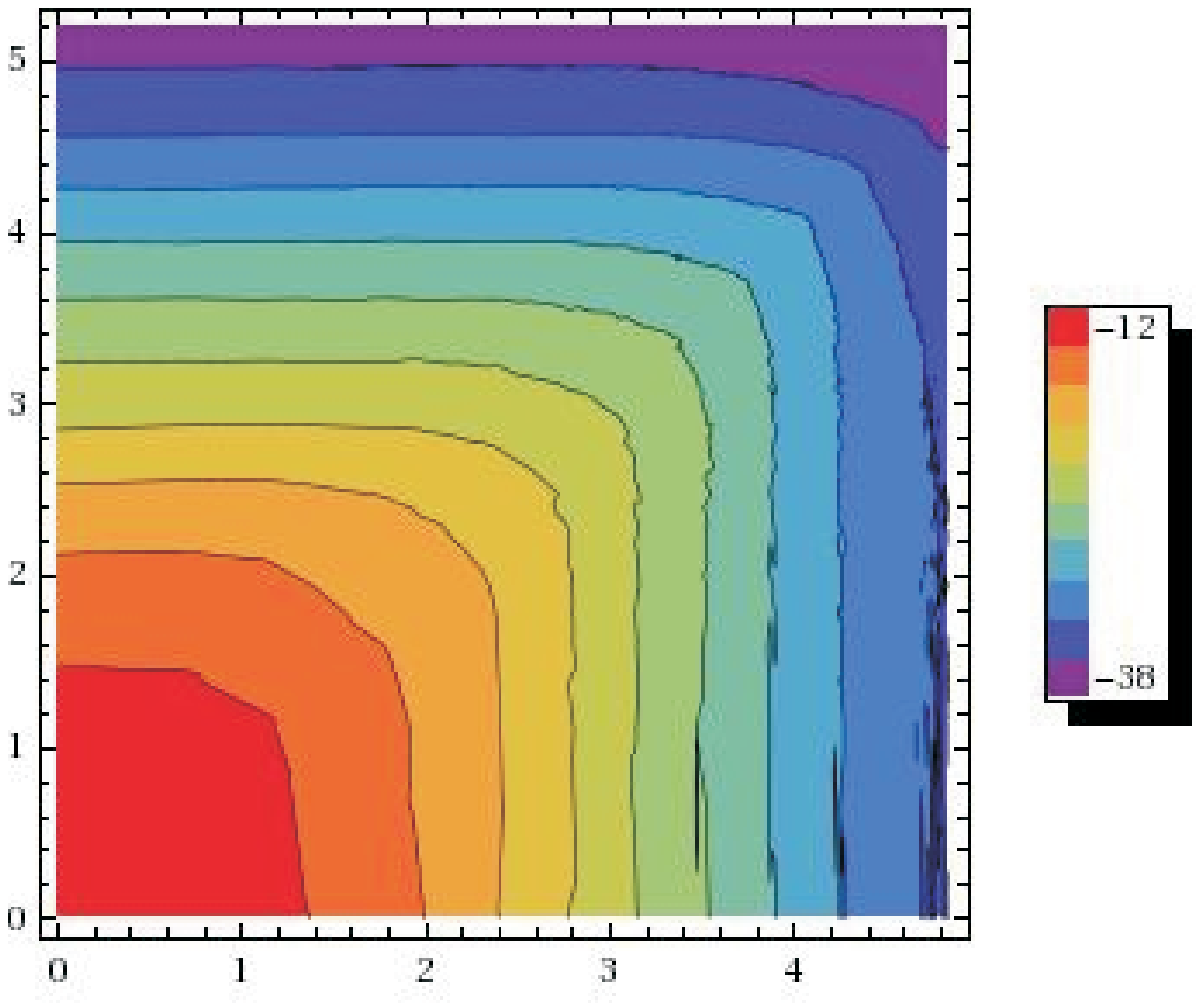}
  \end{picture}
  \label{PShear}
}
\subfigure[PComp]{
\begin{picture}(1000,837)(-100,-100)
\put(-60,330){\rotatebox{90}{$\ln k_{\perp}$}}
\put(360,-60){$\ln k_{\parallel}$}
    \includegraphics[width=900\unitlength]{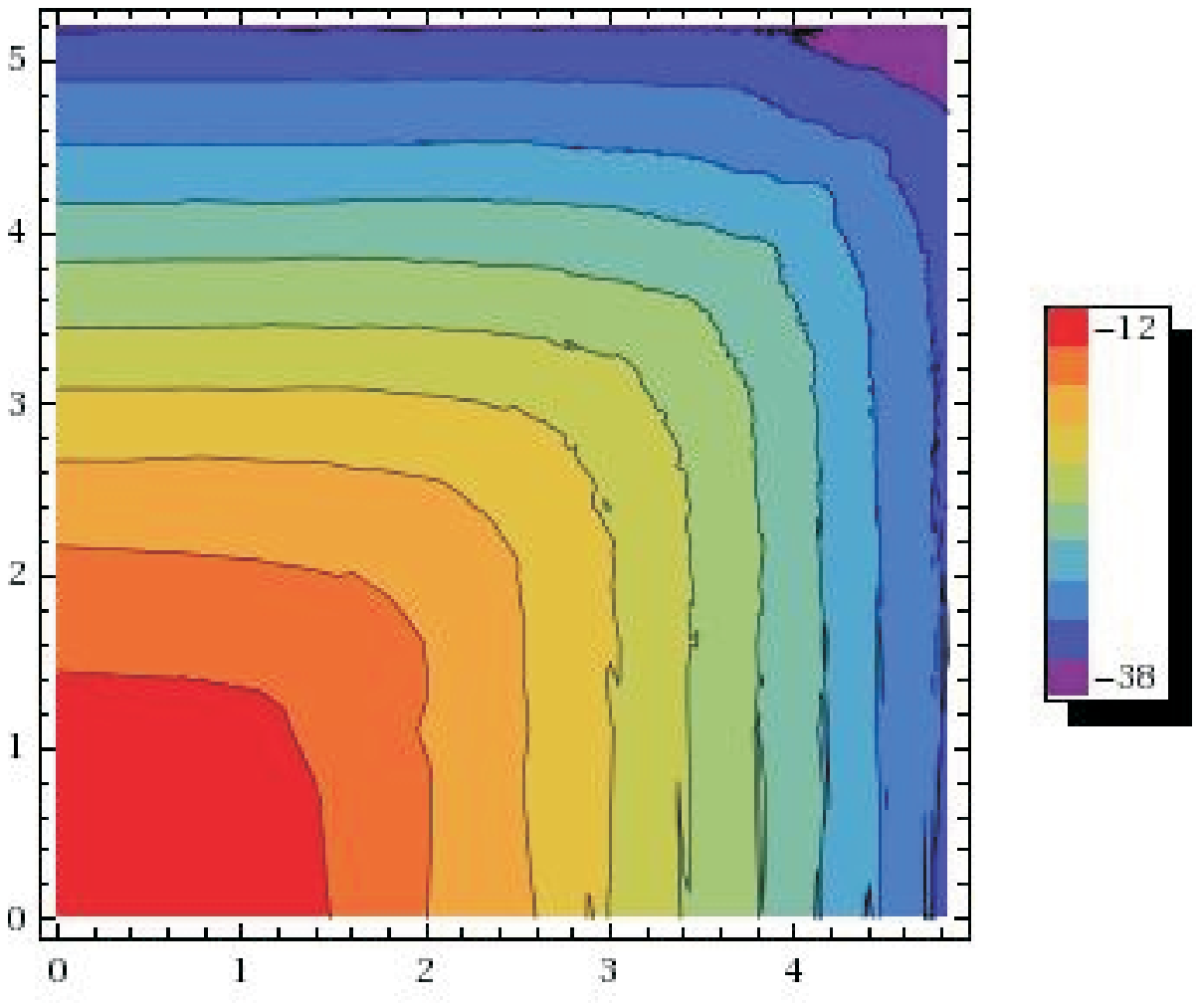}
  \end{picture}
  \label{PComp}
}
\end{center}
\caption{Figs. \ref{PB}, \ref{PK}, \ref{PShear} and \ref{PComp} show the resulting two dimensional spectra of the magnetic field, the velocity field, the incompressible component of the velocity field and the compressible component of the velocity field as contour plots. Apparently all those spectra are highly homogenous or have only a slightly perpendicular preferred direction, i.e. the contour lines are elongated along the $k_\perp$ axis.}
\label{fig:rundplots}
\end {figure}

Additionally it can be seen that the simulated spectrum has a distinct inertial range (Fig. \ref{fig:heli_spek}) and apparently also the correct energy of the fluctuating component. We show the fluctuation energy for our simulations whereas for the solar wind measurements only the perpendicular component of the spectrum was measured. The slab component has been estimated to be about 20 percent which fits well for  several theoretical approaches \citep[cf. eg.][]{bieber96}. This is however not based on measurements, as the decomposition is based on modeling rather than observing. In our simulations based on the solar wind plasma parameters we were not able to find evidence for a strong anisotropy of the spectrum, however (as can be seen in Fig. \ref{fig:rundplots}). With a plasma $\beta$ of 0.85 a strong anisotropy is not to be expected anyway.

The timestep in our simulation is determined by fastest the particles in the simulation domain. This limit is computed in the following section.

\subsubsection{Time step accuracy}
\begin {figure}
\begin{center}
  \includegraphics[width=0.8\textwidth]{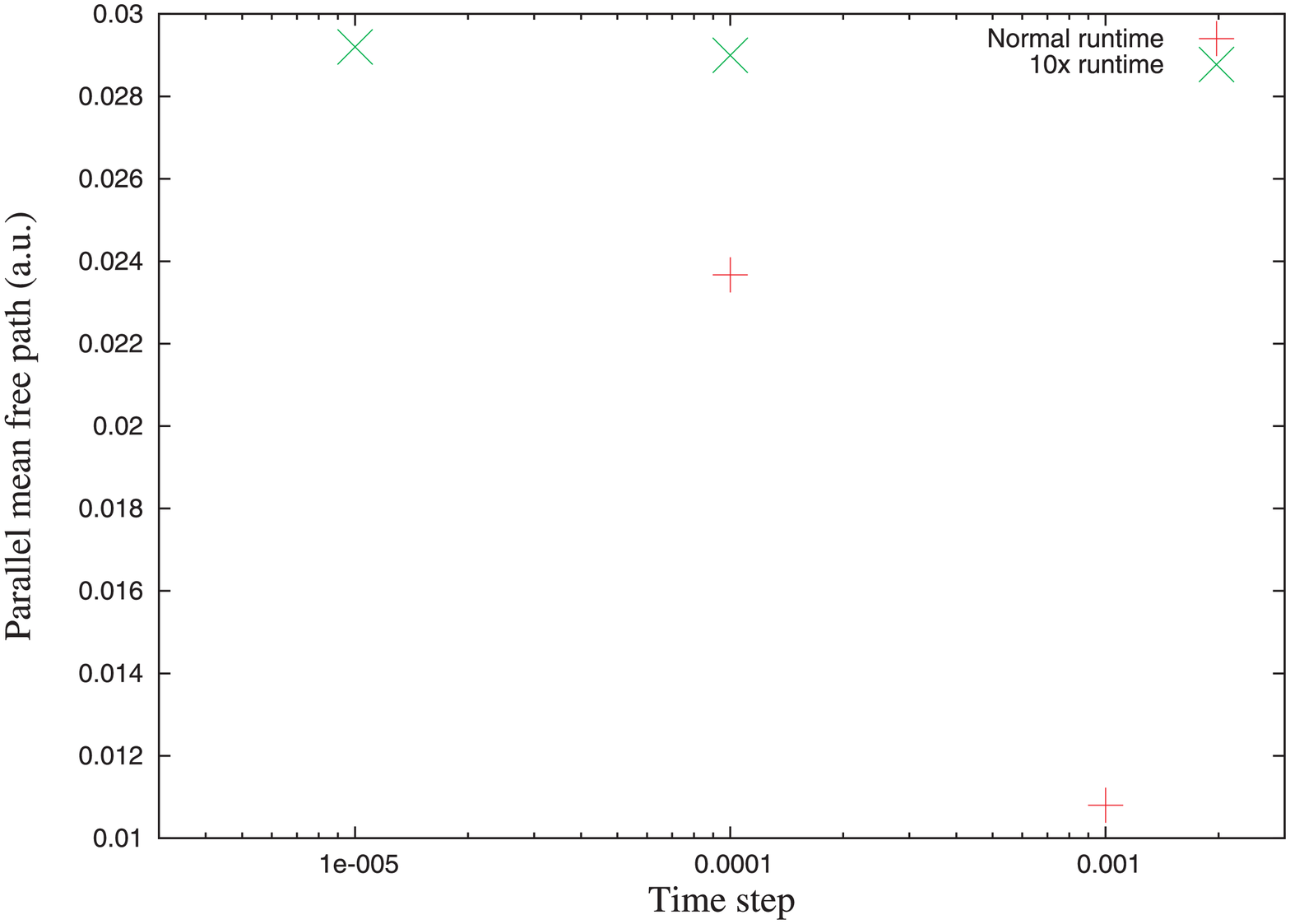}
\end{center}
\caption{This Figure shows the mean free path as a function of  time step sizes. The $10\times$ runtime runs have been simulated longer to achieve better convergence. The difference between the simulation with  $\Delta t = 10^{-4}$  and the one with $\Delta t = 10^{-5}$ is already very small. From this we concluded that a  time step  $\Delta t = 10^{-4}$ is suited for our purpose.
}
\label{fig:Spektrum_grid}
\end {figure}

An upper limit for the time step can be evaluated by comparing the mean free paths computed for two simulations after the same physical time but with a different time step size. This analysis is shown in Fig. \ref{fig:Spektrum_grid}. Since a time step size of $10^{-4}$ gives converged results, the following simulations have been run with a time step of $10^{-4}$.

\subsubsection{Influence of the particle number on the accuracy of the computation of the Fokker-Planck-coefficients}

The last numerical input parameter for the particle simulation is the number of particles within a simulation. As we want to sample all realizations of the spectrum we need to have a sufficiently high number of particles to sample all uncorrelated structures. Apart from that we need good statistics in $\mu$-space to calculate the mean free path correctly. If for instance one would integrate $D_{\mu\mu}(\mu)$ given in Fig. \ref{fig:dmm} to calculate the parallel mean free path one needs very small bin size as $D_{\mu\mu}(\mu)$ has a pronounced $\mu$-dependence.\\
The speed of convergence is essentially dependent on the particle number per bin. This has of course a serious impact on the minimum particle number.

\begin {figure}
\begin{center}
\setlength{\unitlength}{0.00045\textwidth}
  \includegraphics[width=0.5\textwidth]{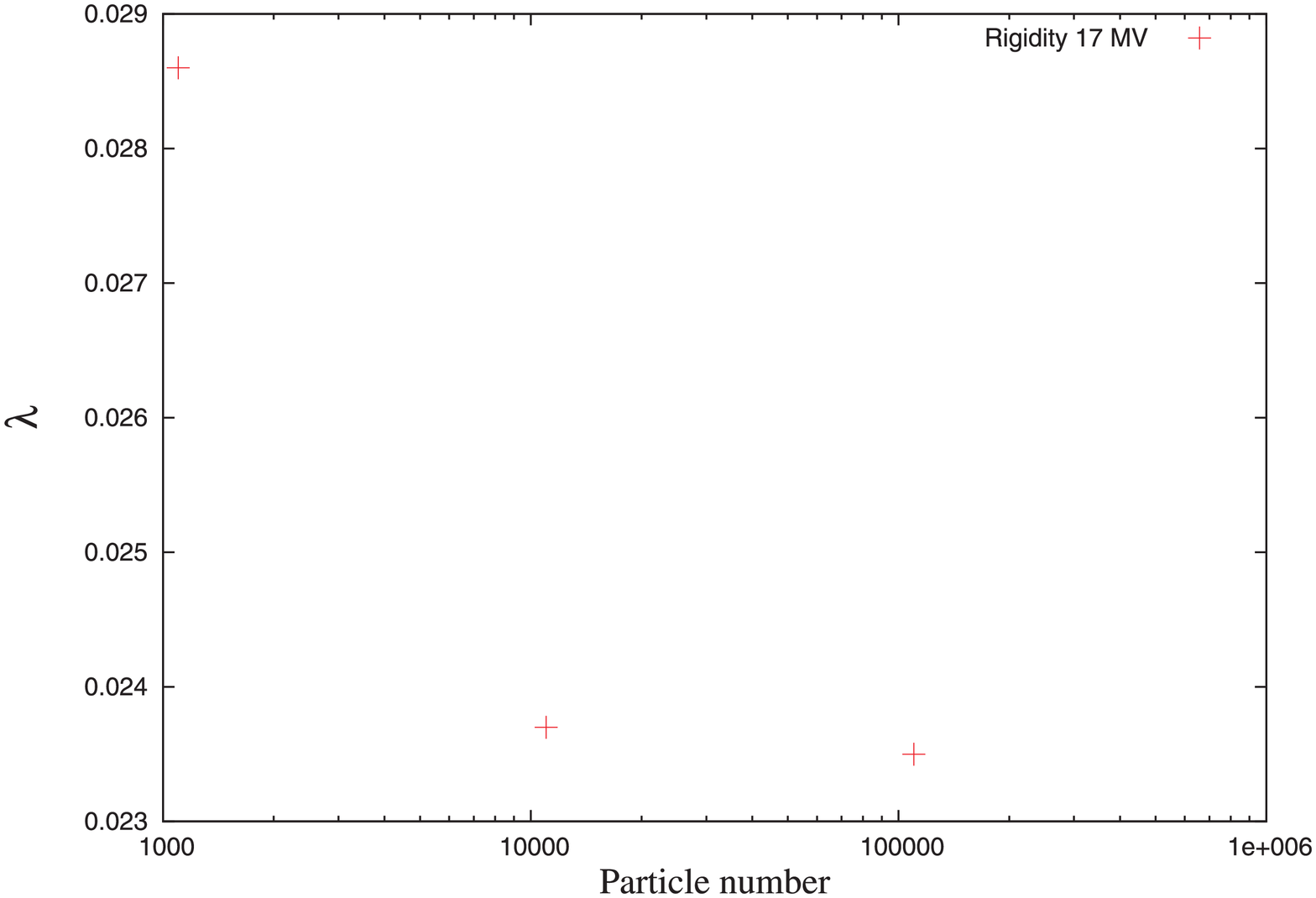}
\end{center}
\caption{The parallel mean free path as a function of  the test particle number $N$. Results are shown for
  1100, 11000 and 110000 particles. }
\label{fig:stabil_pn}
\end {figure}

For this test we compute the mean free path for several different runs - in this case with different numbers of particles. In this context we present simulations with 1100, 11000 and 110000 particles. The results can be seen in Fig. \ref{fig:stabil_pn}. Apparently the results of the two runs with 11000 and 110000 particles are quite close. A simulation with only 1100 particles is however unacceptable.

It should be mentioned that the results of this simulation depend very much on the strength of the turbulence. For small turbulence amplitudes the particles are only slightly scattered. This results in very small values of $D_{\mu\mu}$. As one has to integrate over $1/D_{\mu\mu}$ to find the parallel mean free path (see equation \ref{eq:mfp}) the mean free path (mfp) depends strongly on the minimum of the $D_{\mu\mu}$, which is a known problem in QLT. A slight difference for a bin with low $D_{\mu\mu}$ may result in very different values for the parallel mean free path. In that case we need good statistics for every bin, as overall convergence is not sufficient. In our final simulations the amplitude in single modes is rather high. So we are able to work with a quite small number of particles as small differences within a bin (which might occur if convergence isn't perfectly reached yet) won't have a significant effect on the resulting mean free path. In the following simulation we worked with 110000 particles as for both 11000 and 110000 particles the main numerical costs are due to the evolution of the MHD-fields. So this additional accuracy is not expensive in this case. Arbitrary high particle numbers, however, result in excessive post-processing costs.

\subsection{Simulation of the mean free path}

\begin {figure}
\begin{center}
\setlength{\unitlength}{0.00045\textwidth}
  \includegraphics[width=0.5\textwidth]{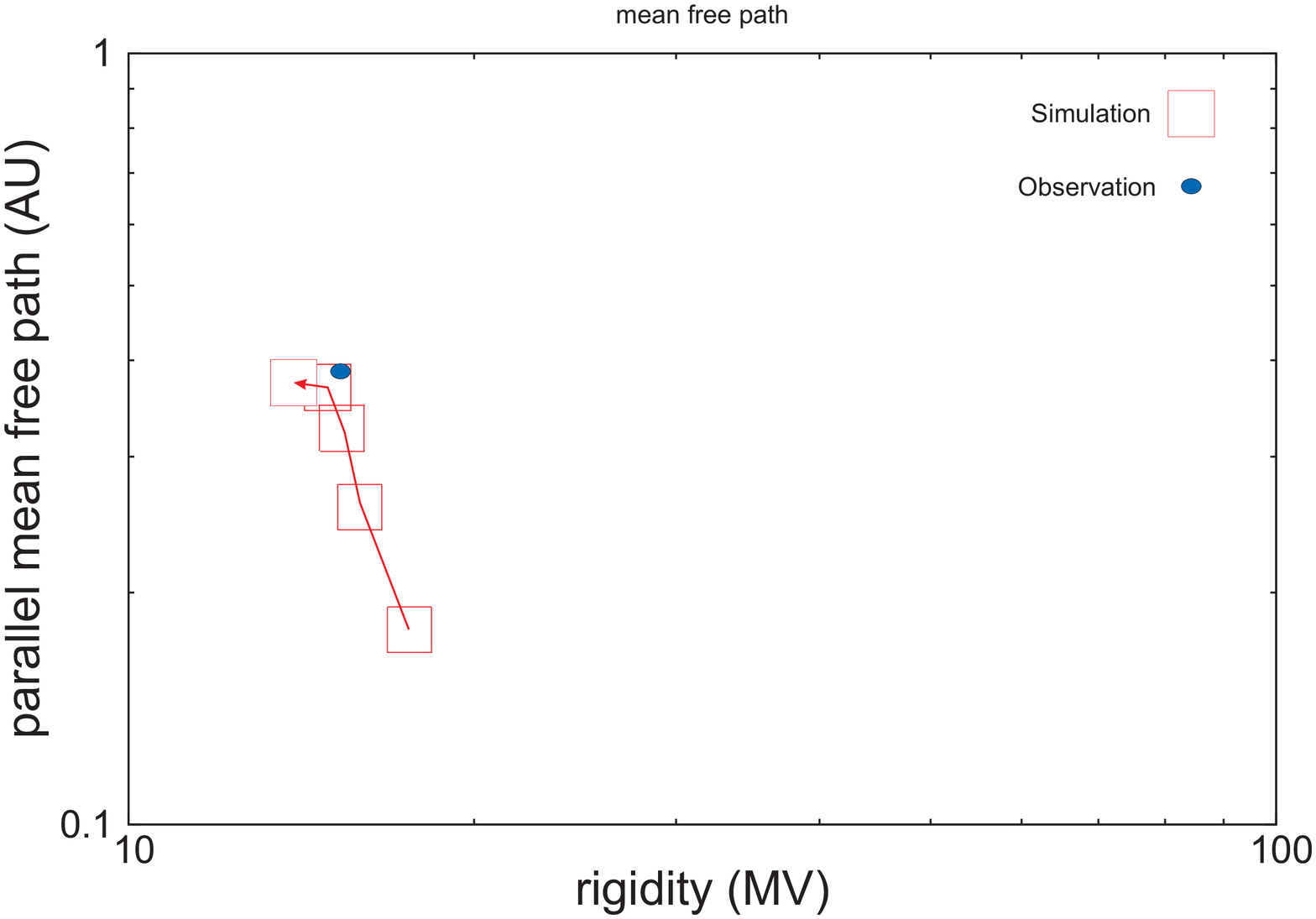}
\end{center}
\caption{The mean free path computed from the numerical model for different, consecutive simulation times. Apparently the mean free path is consistent with the experimental value (see Fig.  \ref{fig:mfwlExp}). The shift in rigidity is due to the energy loss of the particles. The points are given at times $t=0, 70, 140, 210, 280 \Omega_i^{-1}$. Further simulations up to $t=700 \Omega_i^{-1}$ show only minor changes in the mfp.}
\label{mfp}
\end {figure}

Up to this point we determined the conditions for a physically relevant simulation of the parallel mean free path for the event of the period 1996 July 9 18:00 UT to July 10 12:00 UT measured by the Wind 3DP spacecraft. Bearing this discussion in mind we now present our simulation results gained by a simulation with 110000 particles initialised with an isotropic distribution in $\mu$ and a rigidity of 17 MV. The simulation was performed on a $256^3$ grid with a time step size of $10^{-4}$. The mean free path is calculated on the basis of Eqs. \ref{eq:dmm} to \ref{eq:mfp}. The results are shown in Fig.  \ref{mfp} and have to be compared with the experimental data in Fig. \ref{fig:mfwlExp}. Simulations were run for several gyration periods.

The different values for the mean free path in Fig. \ref{mfp} result from the same simulation with an initial particle rigidity of 17 MV. Particle tracks were saved to disk in equidistant time intervals. It can be seen that the simulated mean free path converges to the experimental value, but also that the particles lose energy during the simulation. This effect can be explained by the Landau resonance, which yields an energy loss for suprathermal particles. The fact that this process is present for almost all superthermal particle energies is a strong hint that we are indeed dealing with the Landau resonance. The plasma-beam-system is essentially unstable, but since the energetic particles are test particles the back reaction is not present.

\begin {figure}
\begin{center}
\setlength{\unitlength}{0.00045\textwidth}
\begin{picture}(1000,744)(-100,-100)
  \put(-60,310){\rotatebox{90}{$\Delta\mu$}}
  \put(470,-60){$\mu$}
  \includegraphics[width=900\unitlength]{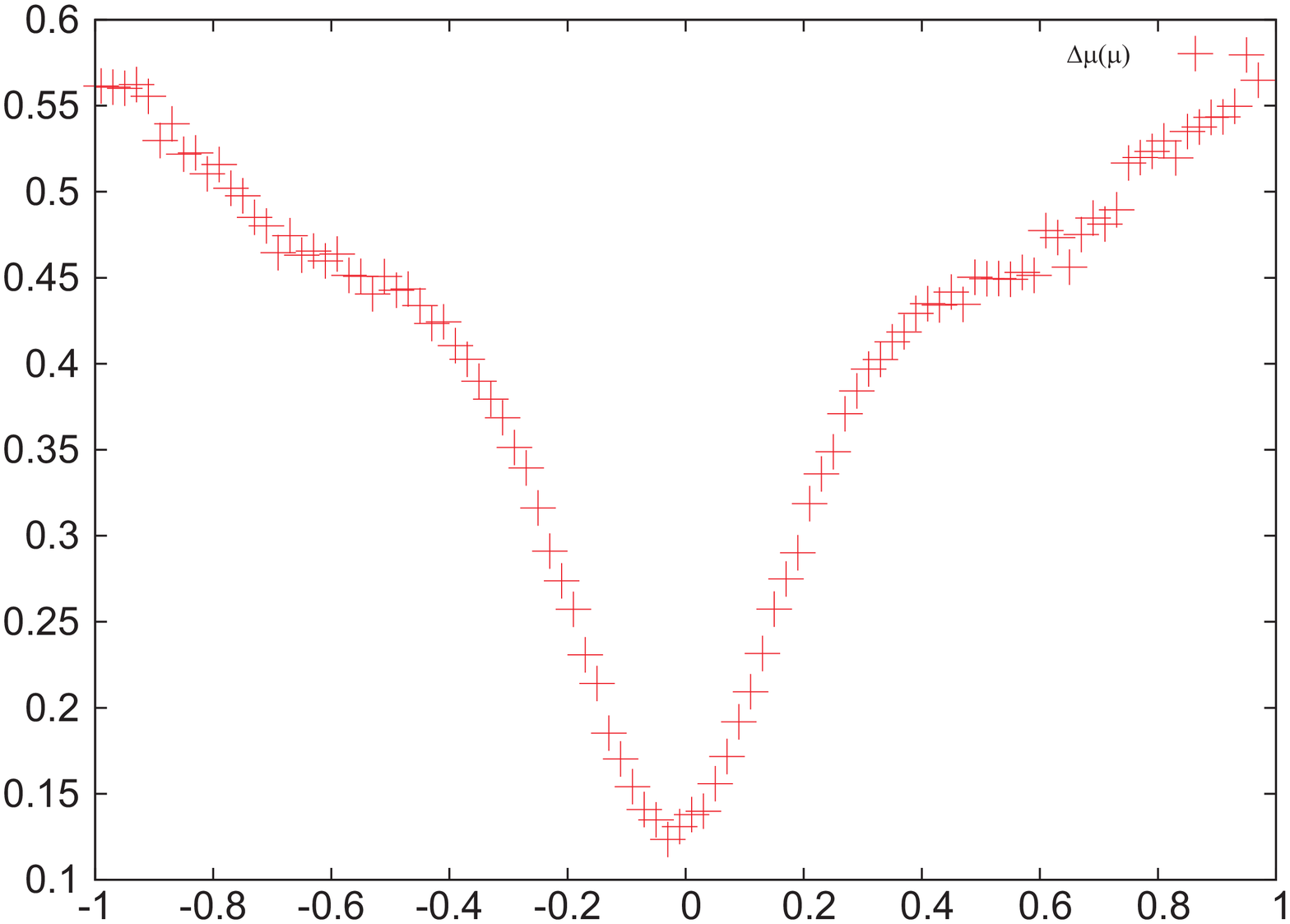}
\end{picture}
\end{center}
\caption{$\Delta\mu(\mu)$ for our numerical model with $\Delta\mu=\mu_{max}-\mu_{min}$. We find a minimum at $\mu=0$. All particles have been binned in  $\mu$-space and the average value of $\Delta \mu$ has been computed for all particles  within a bin.}
\label{streu_dm}
\end {figure}

Fig. \ref{streu_dm} shows the averaged change of the pitch angle for the particles as a function of $\mu$. For this the difference of the maximal and minimal value of $\mu$ during the whole simulation has been calculated for each particle. The particles have been binned in $\mu$-space (initial value of $\mu$) and the average value of $\Delta \mu$ of all particles within a bin has been plotted in Fig. \ref{streu_dm}.

Apparently there is only slight scattering of the particles near $\mu=0$  whereas pitch angle scattering near $\mu=\pm 1$ is significant. Thus, we conclude that the main prerequisite of the QLT is not fulfilled under these conditions:
Particles are scattered so strongly that they may leave the field lines. While this seems to exclude a description via the QLT, it also raises some questions regarding the methods to analyze particle transport. One is related to the theoretical calculation of the pitch-angle diffusion coefficient and the other is related to the data analysis of satellite data. In both cases it is assumed that the diffusion approximation holds and particle diffusion is described by the Fokker-Planck equation. As soon as we deviate from the assumptions of QLT both assumptions are only low-scattering limits of reality. Therefore the calculation of a $D_{\mu\mu}$ is possible but the applicability of the FP equation is limited. This problem exists in the experiment and in theory. Any deviation from this approximation yields the same error in both methods. Therefore theory and observation are comparable even when leaving the realm of QLT.

Two additional aspects of non-QLT effects which are specific to dynamic turbulence: The simpler one is the effect of the electric field, which plays a role when the particle's parallel velocity is smaller or comparable to the wavespeed. This effect has been discussed already, but should not play a major role for $\vert\mu\vert \gg 0$. For particles propagating almost perpendicular to the magnetic field an effect might be visible. The second effect of dynamic turbulence is caused by the correlation function. This has been discussed already as early as \citet{volk1973}. The correlation function in time has a direct influence on the diffusion and is modeled here directly. Unfortunately it is hard to disentangle the effect of temporal evolution, since the correlation function would have to be determined along the particle's path. A quantitative estimate is left for future work.

\subsection{Evolution of particle distribution function}

Within the simulation we also monitor the evolution of particle spectra. Since we are making no quasilinearity assumptions, particles are actually changing pitch angles and may also leave their magnetic field lines. In our present analysis we do not distinguish the effects of field line wandering and cross field diffusion. Testing this would require more sophisticated methods, which will be applied in forthcoming papers.

Our first study was conducted using an isotropic initial particle distribution. This may not be physically motivated, but is very usefeul when studying diffusion for all pitch angles at the same time. It is assumed that pitch angle diffusion may lead to a steady state independent of the initial conditions after sufficiently long time.

\begin{figure}
  \begin{center}
    \includegraphics[width=0.45\textwidth]{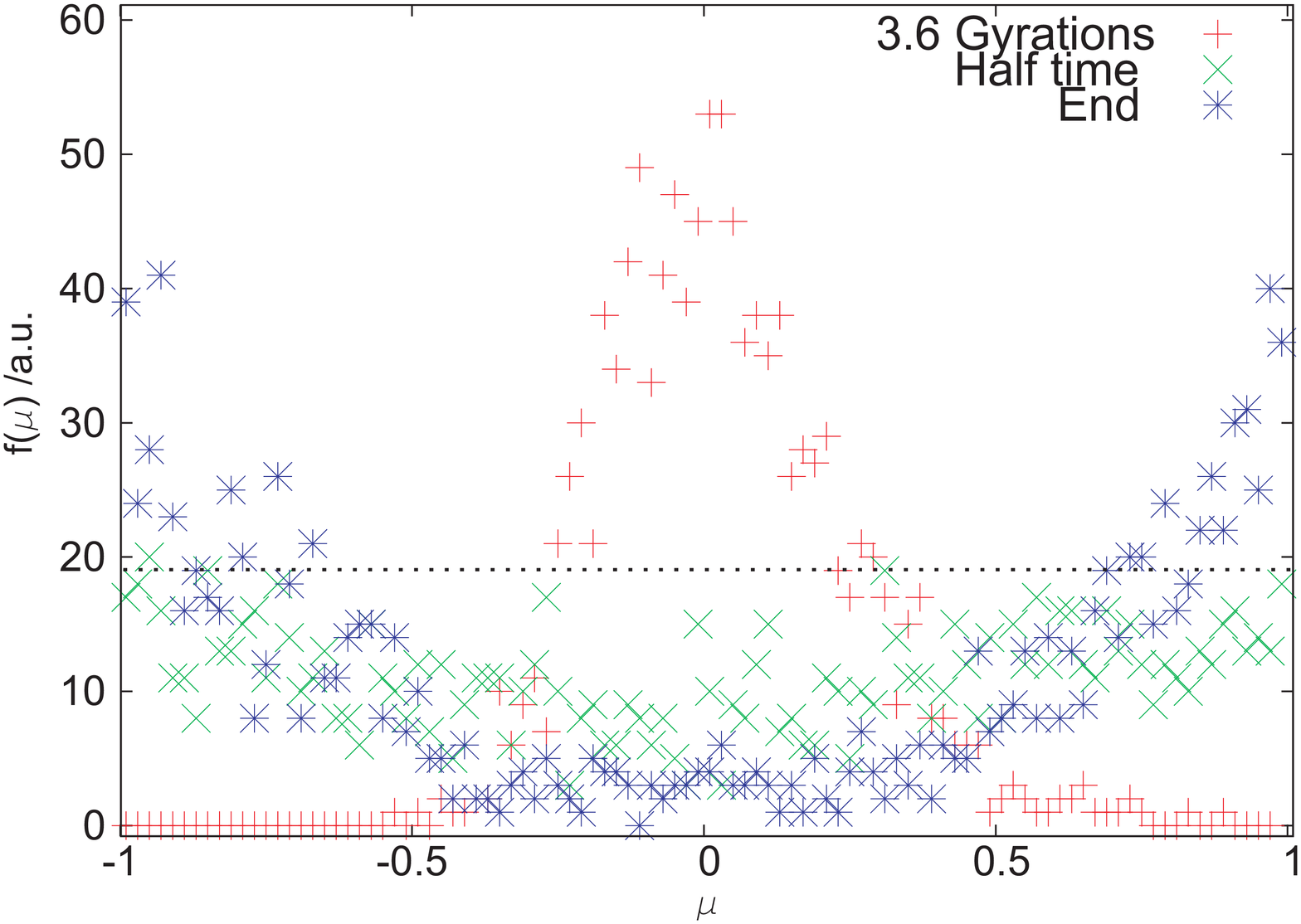}
    \caption{Evolution of the particle distribution starting with an isotropic distribution (shown as dashed line). The distribution is plotted at physical times 63 s, 450 s and 900 s.}
      \label{fig:mudistiso}
        \end{center}
\end{figure}

\begin{figure}
  \begin{center}
    \includegraphics[width=0.45\textwidth]{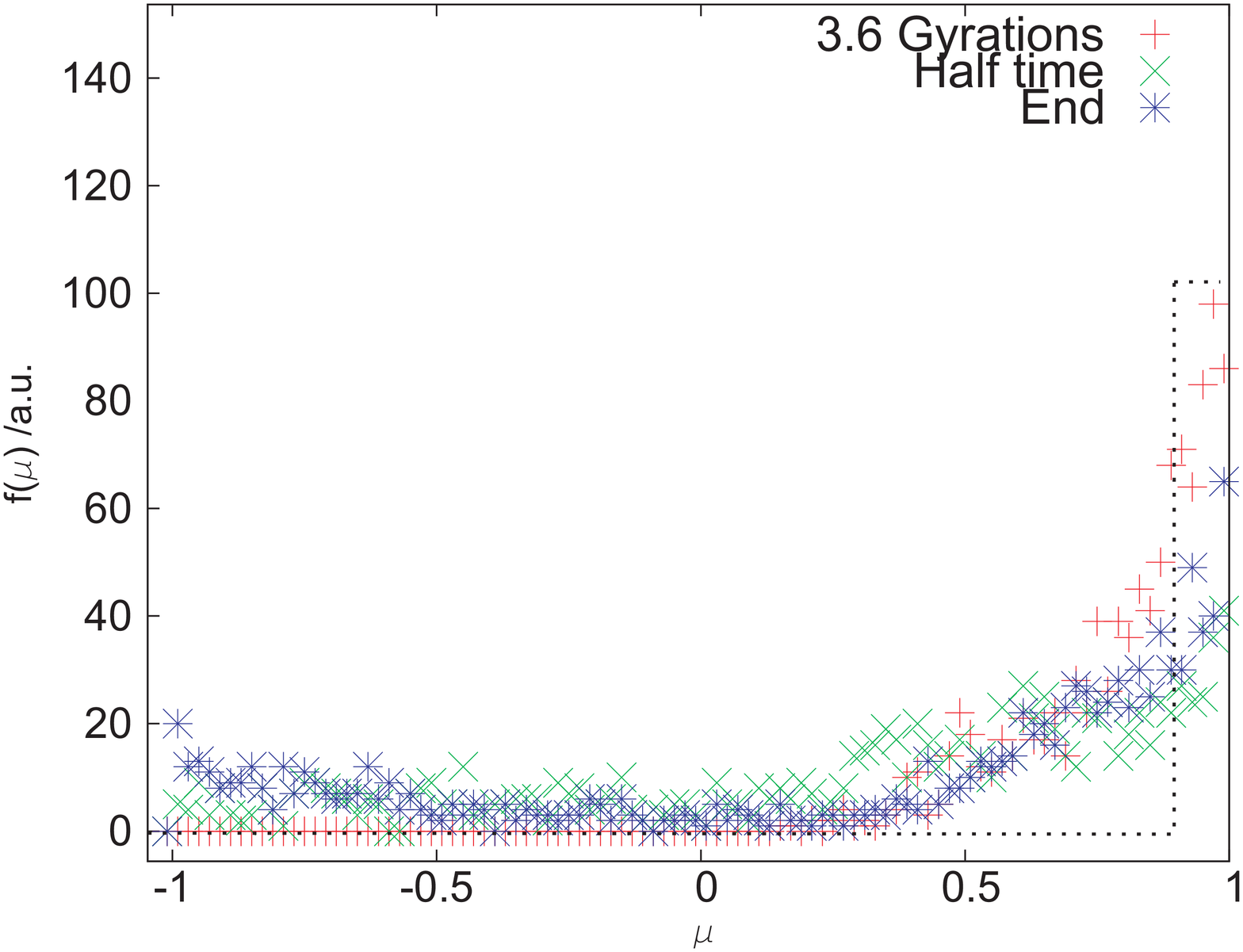}
    \caption{Evolution of the particle distribution starting with a particle distribution between $\mu=0.8$ and $\mu=1$ (shown as dashed line). The distribution is plotted at physical times 63 s, 450 s and 900 s.}
  \label{fig:mudistpar}

  \end{center}
\end{figure}

In Fig. \ref{fig:mudistiso} we  show the evolution from the isotropic initial distribution, while Fig. \ref{fig:mudistpar} shows the same evolution, but for a particle distribution which initially is peaked along the mean field. In both cases the evolution is rather fast in gyrotimescales and shows similar (and not unexpected) behavior: particles align to the magnetic field lines.

\subsection{Pitch-angle diffusion coefficient}

\begin{figure}
  \begin{center}
    \includegraphics[width=0.45\textwidth]{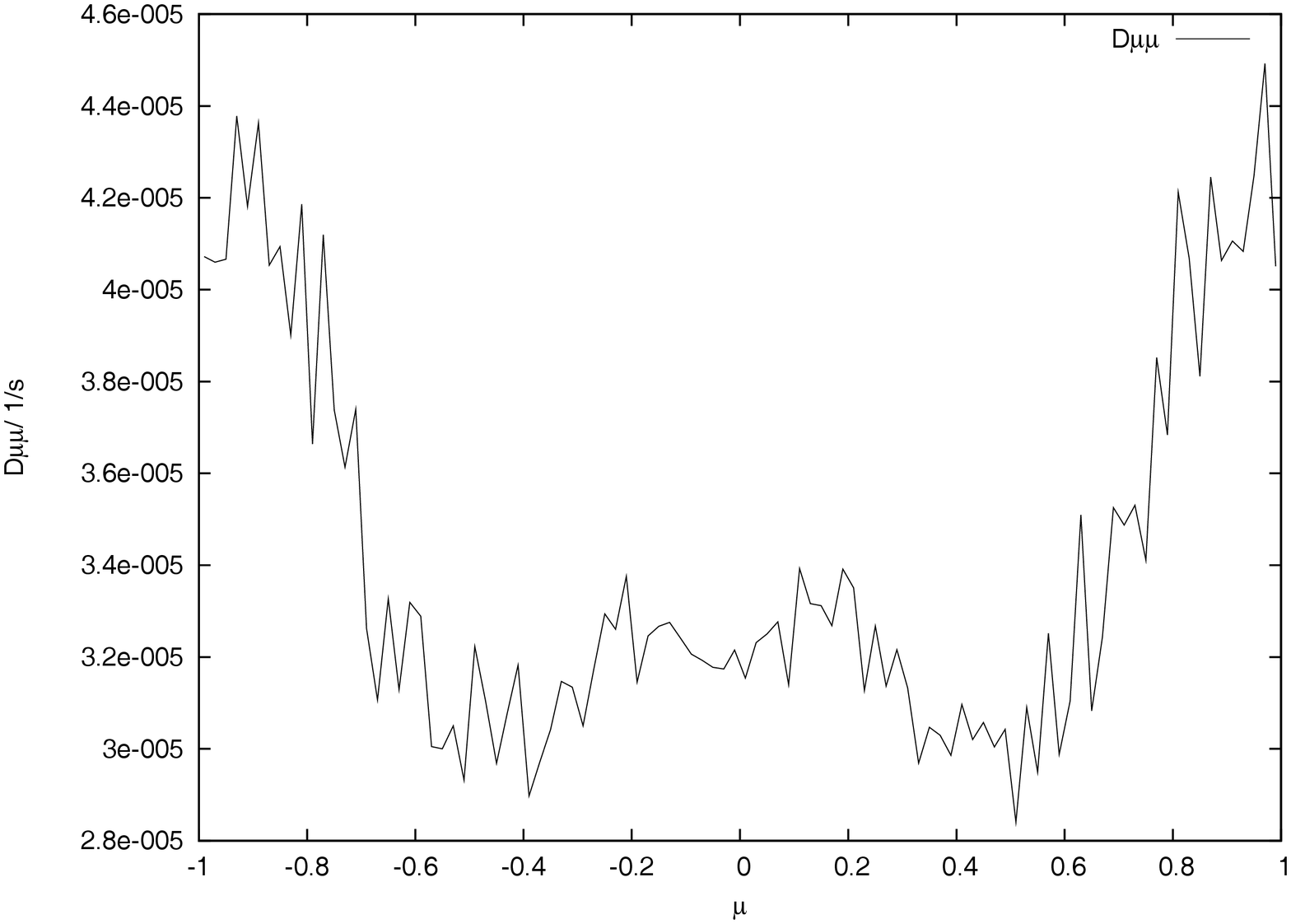}
    \caption{The pitch-angle diffusion coefficient for protons in the described simulation setting.}
    \label{fig:dmm}
  \end{center}
\end{figure}

It has already been noted that the calculation of the diffusion coefficient itself is rather tedious and that it should not be compared with observations. Nevertheless we show one of the diffusion coefficients in Fig. \ref{fig:dmm}. It can be seen that the diffusion coefficient shows in first order the symmetric double peak structure. But there is also  a clear non-zero contribution at $\mu=0$.

Additionally one has to consider the contribution at $\vert\mu\vert=1$. Fig. \ref{fig:dmm} shows maxima close to 1, but particle statistics do not allow to make a clear prediction of a QLT violating result here. As the particle distribution evolves towards a distribution aligned parallel to the field do not predict extreme scattering away from those points. The comparison with results from \citet{2011ASTRA...7...21S} is rather difficult as the spectrum there is limited to one or two points in the spectrum. The soft spectrum at higher $k$ may lead to the maxima close to 1.

\section{Discussion and Conclusions}

In this paper we analysed the transport of energetic particles in an MHD plasma with heliospheric parameters. We verified that the numerical framework can reproduce the resonant wave-particle interaction which is the basis for energetic particle transport. The parallel mean free path which results from this simulation is consistent with satellite observations.

From the numerical point of view it has been shown that moderate grid resolutions are already sufficient to provide good results for the transport parameters. This model is, however, limited to local transport not taking into account the large scale gradients of the solar wind plasma from the Sun to 1 AU. This is a further step to be taken. With our investigations for the heliosphere we were able to proof that our simulations are able to model SEP transport on the basis of resonant wave particle interactions in turbulent MHD fields. As we believe that the transport of CRs in the ISM works in a similar way our results should be transferable to this scenario as well. For the case of ISM turbulence a compressible driving should be more adequate because the turbulence is mainly generated by shocks of SN and SNR (as a incompressible cascade is proposed the driving may depend on the length scale).

The key point in this study is not to show that we are able to reproduce the mean free path which has been done already in previous studies \citep[eg.][]{2003ApJ...589.1027D,1969lhea.conf..111R}, but to shed light on the physical processes in the background. Here one main finding is, that under given conditions QLT is not necessarily the best choice for the mathematical description. Particles scatter stronger on the turbulent fluctuations than QLT would allow. Especially the scattering through $\mu=0$ is not zero, as QLT predicts. Analytic theories have to be modified likewise (see e.g., \citealt{2008ApJ...685L.165T}).
As far as the current simulations are concerned the wave-particle resonance has to be considered as the main transport process.

A critical analysis shows that we have many assumptions in common with \citet{2003ApJ...589.1027D}, but here we made some additional assumptions with regard to the physical background. The most significant ones are the compressionability and anisotropy of the spectrum. Further simulations especially with incompressible spectra have to shed more light on this issue.

In this manuscript we have outlined the possibilities of our ansatz and have proven the applicability to observational data. The next step in its application should of course be the application to a wider range of data to find a general behaviour of transport.

\section*{Acknowledgements}
      MW acknowledges support by Graduiertenkolleg 1147 and FS acknowledges support by
      \emph{Deut\-sche For\-schungs\-ge\-mein\-schaft, DFG\/} project
      number Sp~1124/3.

\bibliographystyle{plainnat}
\bibliography{paperbib}

\end{document}